\documentclass[longauth,twocolumns]{aa}
\usepackage{graphicx}
\usepackage{natbib}
\usepackage{amstext}
\usepackage{color}
\usepackage{txfonts}

\begin{document}

\title{
Near-Infrared interferometry of $\eta$ Carinae with
high spatial and spectral resolution using the VLTI and the AMBER instrument\thanks{
Based on observations made with ESO telescopes at Paranal Observatory 
under programme ID 074.A-9025(A).}
}


\author{G.~Weigelt\inst{1} 
  \and S.~Kraus\inst{1}
  \and T.~Driebe\inst{1}
  \and R.G.~Petrov\inst{4}
  \and K.-H.~Hofmann\inst{1}
  \and F.~Millour\inst{2,4}
  \and O.~Chesneau\inst{3}
  \and D.~Schertl\inst{1}
  \and F.~Malbet\inst{2}
  \and J.D.~Hillier\inst{18}
  \and T.~Gull\inst{17}
  \and K.~Davidson\inst{16}
  \and A.~Domiciano de Souza\inst{3,4}
  \and P.~Antonelli\inst{3}
  \and U.~Beckmann\inst{1}
  \and Y.~Bresson\inst{3}
  \and A.~Chelli\inst{2}
  \and M.~Dugu\'e\inst{3}
  \and G.~Duvert\inst{2}
  \and L.~Gl\"uck\inst{2}
  \and P.~Kern\inst{2}
  \and S.~Lagarde\inst{3}
  \and E.~Le~Coarer\inst{2}
  \and F.~Lisi\inst{5}
  \and K.~Perraut\inst{2}
  \and P.~Puget\inst{2}
  \and S.~Robbe-Dubois\inst{4}
  \and A.~Roussel\inst{3}
  \and G.~Zins\inst{2}
  \and M.~Accardo\inst{5}
  \and B.~Acke\inst{2,13}
  \and K.~Agabi\inst{4}
  \and B.~Arezki\inst{2}
  \and E.~Altariba\inst{4}
  \and C.~Baffa\inst{5}
  \and J.~Behrend\inst{1}
  \and T.~Bl\"ocker\inst{1}
  \and S.~Bonhomme\inst{3}
  \and S.~Busoni\inst{5}
  \and F.~Cassaing\inst{6}
  \and J.-M.~Clausse\inst{3}
  \and J.~Colin\inst{3}
  \and C.~Connot\inst{1}
  \and A.~Delboulb\'e\inst{2}
  \and P.~Feautrier\inst{2}
  \and D.~Ferruzzi\inst{5}
  \and T.~Forveille\inst{2}
  \and E.~Fossat\inst{4}
  \and R.~Foy\inst{7}
  \and D.~Fraix-Burnet\inst{2}
  \and A.~Gallardo\inst{2}
  \and S.~Gennari\inst{5}
  \and A.~Glentzlin\inst{3}
  \and E.~Giani\inst{5}
  \and C.~Gil\inst{2}
  \and M.~Heiden\inst{1}
  \and M.~Heininger\inst{1}
  \and O.~Hernandez\inst{2}
  \and D.~Kamm\inst{3}
  \and M.~Kiekebusch\inst{11}
  \and D.~Le Contel\inst{3}
  \and J.-M.~Le Contel\inst{3}
  \and B.~Lopez\inst{3}
  \and Y.~Magnard\inst{2}
  \and A.~Marconi\inst{5}
  \and G.~Mars\inst{3}
  \and G.~Martinot-Lagarde\inst{8,14}
  \and P.~Mathias\inst{3}
  \and J.-L.~Monin\inst{2}
  \and D.~Mouillet\inst{2,15}
  \and D.~Mourard\inst{3}
  \and P.~M\`ege\inst{2}
  \and E.~Nussbaum\inst{1}
  \and K.~Ohnaka\inst{1}
  \and J.~Pacheco\inst{3}
  \and C.~Perrier\inst{2}
  \and Y.~Rabbia\inst{3}
  \and F.~Rantakyr\"o\inst{11}
  \and S.~Rebattu\inst{3}
  \and F.~Reynaud\inst{9}
  \and A.~Richichi\inst{10}
  \and M.~Sacchettini\inst{2}
  \and P.~Salinari\inst{5}
  \and M.~Sch\"oller\inst{11}
  \and W.~Solscheid\inst{1}
  \and P.~Stee\inst{3}
  \and P.~Stefanini\inst{5}
  \and M.~Tallon\inst{7}
  \and I.~Tallon-Bosc\inst{7}
  \and E.~Tatulli\inst{2}
  \and D.~Tasso\inst{3}
  \and L.~Testi\inst{5}
  \and F.~Vakili\inst{4}
  \and J.-C.~Valtier\inst{3}
  \and M.~Vannier\inst{4,11}
  \and N.~Ventura\inst{2}
  \and 0.~von~der~L\"uhe\inst{20}
  \and K.~Weis\inst{19}
  \and M.~Wittkowski\inst{10}
}

\offprints{G.~Weigelt\\ email: \texttt{weigelt@mpifr-bonn.mpg.de}}

\institute{
  Max-Planck-Institut f\"ur Radioastronomie, Auf dem H\"ugel 69,
  D-53121 Bonn, Germany
  \and Laboratoire d'Astrophysique de Grenoble, UMR 5571 Universit\'e Joseph
  Fourier/CNRS, BP 53, F-38041 Grenoble Cedex 9, France
  \and Laboratoire Gemini, UMR 6203 Observatoire de la C\^ote
  d'Azur/CNRS, Avenue Copernic, 06130 Grasse, France
  \and Laboratoire Universitaire d'Astrophysique de Nice, UMR 6525
  Universit\'e de Nice/CNRS, Parc Valrose, F-06108 Nice cedex 2, France
  \and INAF-Osservatorio Astrofisico di Arcetri, Istituto Nazionale di
  Astrofisica, Largo E.~Fermi 5, I-50125 Firenze, Italy
  \and ONERA/DOTA, 29 av de la Division Leclerc, BP 72, F-92322
  Chatillon Cedex, France
  \and Centre de Recherche Astronomique de Lyon, UMR 5574 Universit\'e
  Claude Bernard/CNRS, 9 avenue Charles Andr\'e, F-69561 Saint Genis
  Laval cedex, France
  \and Division Technique INSU/CNRS UPS 855, 1 place Aristide
  Briand, F-92195 Meudon cedex, France
  \and IRCOM, UMR 6615 Universit\'e de Limoges/CNRS, 123 avenue Albert
  Thomas, F-87060 Limoges cedex, France
  \and European Southern Observatory, Karl Schwarzschild Strasse 2,
  D-85748 Garching, Germany
  \and European Southern Observatory, Casilla 19001, Santiago 19,
  Chile
  \and Dipartimento di Fisica, Universit\`a degli Studi di Milano, Via
  Celoria 16, I-20133 Milano, Italy
  \and Instituut voor Sterrenkunde, KULeuven, Celestijnenlaan 200B,
  B-3001 Leuven, Belgium
  \and \emph{Present affiliation:} Observatoire de la Côte d'Azur -
  Calern, 2130 Route de l'Observatoire , F-06460 Caussols, France
  \and \emph{Present affiliation:} Laboratoire Astrophysique de
  Toulouse, UMR 5572 Universit\'e Paul Sabatier/CNRS, BP 826, F-65008
  Tarbes cedex, France
  \and School of Physics and Astronomy, University of Minnesota, 
  116 Church Street SE, Minneapolis, MN 55455, USA
  \and Laboratory for Extraterrestrial Planets and Stellar Astrophysics,
  Goddard Space Flight Center, 20771 Greenbelt, Maryland, USA
  \and Department of Physics and Astronomy, University of Pittsburgh, 
  3941 O'Hara Street, Pittsburgh, PA 15260, USA
  \and Astronomisches Institut, Ruhr-Universit\"at Bochum,
  Universit\"atsstr. 150, D-44780 Bochum, Germany; \& Lise-Meitner fellowship
  \and Kiepenheuer-Institut f\"ur Sonnenphysik, Sch\"oneckstr. 6, D-79104 Freiburg, Germany
}

\date{Received date; accepted date}

\titlerunning{AMBER/VLTI observations of $\eta$~Carinae}

\authorrunning{G.~Weigelt et al.\ }

\abstract
{}
{
We present the first NIR \textit{spectro-interferometry} of the LBV 
\object{$\eta$~Carinae}. The observations were performed with the AMBER instrument of the ESO 
\textit{Very Large Telescope Interferometer} (VLTI) using baselines from 42 to 89~m. The aim 
of this work is to study the wavelength dependence of $\eta$~Car's optically thick wind region 
with a high spatial resolution of 5~mas (11~AU) and \textit{high spectral resolution.}
}
{
The observations were carried out with three 8.2 m Unit Telescopes in the $K$-band.
The raw data are spectrally dispersed interferograms obtained with spectral
resolutions of 1,500 (MR-K mode) and 12,000 (HR-K mode). The MR-K
observations were performed in the wavelength range around both the
\ion{He}{I}\,2.059\,$\mu$m and the Br$\gamma$\,2.166\,$\mu$m emission 
lines, the HR-K observations only in the Br$\gamma$ line region.
}
{
The spectrally dispersed AMBER interferograms allow the investigation of the
\textit{wavelength dependence} of the visibility, differential phase, and
closure phase of $\eta$~Car. In the $K$-band continuum, a diameter of $4.0\pm0.2$~mas 
(Gaussian FWHM, fit range 28--89~m baseline length) was measured for $\eta$~Car's optically 
thick wind region. If we fit Hillier et al.\ (2001) model visibilities to the observed 
AMBER visibilities, we obtain 50\% encircled-energy diameters of 4.2, 6.5 and 9.6~mas 
in the 2.17$\,\mu$m continuum, the \ion{He}{I}, and the Br$\gamma$ emission lines, 
respectively. In the continuum near the Br$\gamma$ line, an elongation along 
a position angle of $120\degr\pm15\degr$ was found, consistent with previous VLTI/VINCI 
measurements by van Boekel et al.\ (2003). We compare the measured visibilities with 
predictions of the radiative transfer model of Hillier et al.\ (2001), finding good agreement.
Furthermore, we discuss the detectability of the hypothetical hot binary companion.
For the interpretation of the non-zero differential and closure phases measured within the 
Br$\gamma$ line, we present a simple geometric model of an inclined, latitude-dependent wind 
zone. Our observations support theoretical models of anisotropic winds from fast-rotating, 
luminous hot stars with enhanced high-velocity mass loss near the polar regions.
}
{}


   \maketitle
%
%
\section{Introduction}\label{sect_intro}

The enigmatic object $\eta$ Car is one of the most luminous and most massive ($M\sim100\,M_\odot$) 
unstable Luminous Blue Variables suffering from an extremly high mass loss \citep{dh97}. Its distance 
is approximately 2300$\pm$100~pc \citep{dh97,davidsonetal01,smi06}. $\eta$~Car, which has been subject 
to a variety of studies over the last few decades, is surrounded by the expanding bipolar Homunculus 
nebula ejected during the Great Eruption in 1843. The inclination of the polar axis of the Homunculus 
nebular with the line-of-sight is $\sim 41\degr$ with the southern pole pointing towards us 
\citep{davidsonetal01,smi06}. The first measurements of structures in the innermost sub-arcsecond 
region of the Homunculus were obtained by speckle-interferometric observations \citep{we86,hw88}. 
These observations revealed a central object (component A) plus three compact and surprisingly bright 
objects (components B, C, and D) at distances ranging from approximately 0.1$^{''}$ to 0.2$^{''}$. 
HST observations of the inner $1^{''}$ region \citep{wei95} provided estimates of the proper motion 
of the speckle objects B, C, and D (velocity $\sim50$ km/s; the low velocity suggests that the speckle 
objects are located within the equatorial plane), and follow-up HST spectroscopy unveiled their unusual 
spectrum \citep{dav95}. The central object (speckle object A) showed broad emission lines, while the 
narrow emission lines came from the speckle objects B, C, and D. Therefore, A is certainly the central 
object while B,C, and D are ejecta. Recent observations of $\eta$ Car by \cite{chesneau05} using NACO 
and VLTI/MIDI revealed a butterfly-shaped dust environment at $3.74\,$ and $4.05\,\mu$m and resolved 
the dusty emission from the individual speckle objects with unprecedented angular resolution in the NIR. 
Chesneau et al.\  also found a large amount of corundum dust peaked $\sim1^{''}$ south-east of the 
central object.

Spectroscopic studies of the Homunculus nebula showed that the stellar wind of $\eta$~Car is 
aspherical and latitude-dependent, and the polar axes of the wind and the Homunculus
appear to be aligned \citep[bipolar wind model;][]{smi03}. Using Balmer line observations obtained 
with HST/STIS, \citet{smi03} found a considerable increase of the wind velocity from the equator to 
the pole and that the wind density is higher in polar direction \citep[parallel to the Homunculus; 
PA  of the axis $\sim$132\degr; ][]{davidsonetal01} than in equatorial direction by a factor of 
$\sim$2. \citet{van03} resolved the optically thick, aspheric wind region with NIR interferometry 
using the VLTI/VINCI instrument. They measured a size of 5 mas (50\% encircled-energy diameter), 
an axis ratio of $1.25\pm0.05$, and a position angle (PA) of the major axis of $134\degr\pm7\degr$, 
and derived a mass-loss rate of $1.6\times10^{-3}\,M_\odot/{\rm yr}^{-1}$. The aspheric wind can be 
explained by models for line-driven winds from luminous hot stars rotating near their critical speed 
\citep[e.g.,][]{owo96,owo98}. The models predict a higher wind speed and density along the polar axis 
than in the equatorial plane. In addition, van Boekel et al.\ showed that the $K$ broad-band observations 
obtained with VINCI are in agreement with the predictions from the detailed spectroscopic model by 
\citet{hillier01}. 

The \citet{hillier01,hillier06} model was developed to explain STIS HST spectra. The luminosity of the 
primary ($5 \times 10^6\,L_\odot$) was set by observed IR fluxes \citep[see discussion by ][]{dh97}
and the known distance of 2.3~kpc to $\eta$~Car. Any contribution to the IR fluxes by a binary companion 
was neglected.  Modeling of the spectra was undertaken using CMFGEN, a non-LTE line blanketed radiative 
transfer developed to model stars with extended outflowing atmospheres \citep{himi98}. For the modeling 
of $\eta$~Carinae, ions of H,  He, C, N, O, Na, Mg, Al, Si, S, Ca, Ti, Cr, Mn, Fe, Ni, and Co were 
included. The mass loss was derived from the strength of the hydrogen lines and their associated electron 
scattering wings. Due to a degeneracy between the mass-loss rate and the He abundance, the H/He helium 
abundance ratio could not be derived, but was set at 5:1 (by number), which is similar to that found by 
\citet{davidson86} from nebula studies. CNO abundances were found to be consistent with those expected 
for full CNO processing. With the exception of Na (which was found to be enhanced by at least a factor of 
2), the adoption of solar abundances for other metal species was found to yield satisfactory fits to the 
STIS spectra. A more recent discussion of the basic model, with particular reference to the UV and outer 
wind, is given by \citet{hillier06}.

Because the wind is optically thick, the models are fairly insensitive to the radius adopted for the 
hydrostatic core (i.e., the radius at which the velocity becomes subsonic). One exception was the 
\ion{He}{I} lines, which decreased in strength as the radius increased and, in general, were very 
sensitive to model details. Additional HST STIS observations show that the \ion{He}{I} lines are 
strongly variable and blue-shifted throughout most of the 5.54-year variability period. These 
observations cannot be explained in the context of a spherical wind model. It now appears likely that 
a large fraction of the \ion{He}{I} line emission originates in the bow shock and an ionization zone, 
associated with the wind-wind interaction zone in a binary system \citep{davidsonetal99,davidson01,
hillier06,nielsen06}. Consequently, the hydrostatic radius derived by \citet{hillier01} is likely to 
be a factor of 2 to 4 too small. Because the wind is so thick, a change in radius will not affect the 
Br$\gamma$ formation region, and it will only have a minor influence on the Br$\gamma$ continuum emitting 
region. If this model is correct, the \ion{He}{I} emission will be strongly asymmetrical and offset 
from the primary star.

A variety of observations suggest that the central source of $\eta$ Car is a binary. \citet{dam96} first 
noticed the 5.5-year periodicity in the spectroscopic changes of this object \citep[see][]{dam97,dam00,
duncan99,ishibashi99,davidsonetal99,davidsonetal00,vangend03,sd04,whitelock04,corcoran05,weis05}.
On the other hand, to date, the binary nature of the central object in $\eta$~Car and its orbital 
parameters are still a matter of debate \citep[see, e.g.,][]{ZWS84,davidson99,davidson01,davidsonetal99,
davidsonetal00,davidsonetal05,ishibashi99,smith00,feast01,ishibashi01,pico02,smi03,martin06}.

The 1997.9 X-ray peak with the subsequent rapid drop to a few-month-long minimum was detected by RXTE 
\citep[see][]{corcoran05}. Then the first spectra with HST/STIS were obtained at 1998.0, demonstrating 
changes in both the central star and the aforementioned speckle objects \citep{davidsonetal99,gull99}.
\citet{pico02} demonstrated that the CHANDRA X-ray spectrum can be explained by the wind-wind collisions 
of the primary star ($\dot{M}=2\times10^{-4}\,M_\odot/{\rm yr}^{-1}$ at 500 km/s) and a hot companion 
($\dot{M}=10^{-5}\,M_\odot/{\rm yr}^{-1}$ at 3,000 km/s). \citet{verner05b} used models calculated with the 
CLOUDY code to demonstrate that during the spectroscopic minimum, the excitation of the speckle objects 
is supported by the primary stellar flux, but that the UV flux of a hot companion consistent with an O7.5V, 
O9I, or early WN star was probably necessary to excite the speckle objects during the broad spectroscopic 
maximum. 

In this paper we present the first spectro-interferometric $K$-band observations of $\eta$ Car obtained 
with the VLTI beam-combiner instrument AMBER with medium and high spectral resolution and in the projected 
baseline range from 28 to 89~m. 

\begin{table*}[t]
\caption{Summary of the AMBER $\eta$~Car observations using the UT2, UT3, and UT4 telescopes.}
\label{tab:observations}
\centering
\begin{tabular}{lll cccccccc}
  \hline\hline
  Date [UT]    &\multicolumn{2}{c}{Time [UT]} &Orbital  &spectral &line within                   &DIT$^d$  &$N_{\eta~Car}^f$&Calibrator &$N_{calib.}^g$   &Calibrator\\
  &Start  &End                                &phase$^e$&mode     &spectral range                &         &                &           &               &uniform disk \\
  &       &                                   &         &         &                              &[ms]     &                &           &               &diameter [mas]\\
  \noalign{\smallskip}
  \hline
  \noalign{\smallskip}
  2004 Dec. 26 & 07:52 & 08:16                & 0.267  &MR-K     & Br$\gamma$                &40                &7,500         &HD\,93030   &5,000 & 0.39$^a$\\
               & 08:19 & 08:32                & 0.267  &MR-K     & \ion{He}{I}~2.059\,$\mu$m &40                &5,000         &HD\,93030   &5,000 & 0.39$^a$\\
  2005 Feb. 25 & 04:33 & 04:43                & 0.298  &MR-K     & Br$\gamma$                &50                &5,000         &HD\,89682   &2,500 & 3.08$^b$\\
               & 04:55 & 05:05                & 0.298  &MR-K     & \ion{He}{I}~2.059\,$\mu$m &50                &5,000         &HD\,89682   &2,500 & 3.08$^b$\\
  2005 Feb. 26 & 08:16 & 08:57                & 0.299  &HR-K     & Br$\gamma$                &82                &7,500         &L\,Car      &2,500 & 2.70$^c$\\
  \noalign{\smallskip}
  \hline
\end{tabular}

\begin{flushleft}
  \hspace{0mm}Notes~--~$^{a}$ Uniform disk (UD) diameter estimated using the method described by~\citet{dyc96}.\\
  \hspace{0mm}$^{b}$ UD diameter taken from the CHARM2 catalog~\citep{ric05}.\\
  \hspace{0mm}$^{c}$ UD diameter of L~Car at the time of the AMBER high-resolution observations
derived from the limb-darkened diameter $d_{\rm LD}=2.80\,$ mas at the L~Car
pulsation phase $\phi$=0.0 \citep{ker04a,ker06} and $d_{\rm UD}/d_{\rm LD}=0.966$ \citep{ker04b}.\\
  \hspace{0mm}$^{d}$ Detector integration time per interferogram.\\
  \hspace{0mm}$^{e}$ The orbital phase was computed assuming a zero point at
  JD 2\,450\,800.0 and a period of 2024~days \citep{corcoran05}.\\
  \hspace{0mm}$^{f}$ Number of $\eta$ Car interferograms.\\
  \hspace{0mm}$^{g}$ Number of calibrator interferograms.\\
\end{flushleft}
\vspace*{-3mm}

\end{table*}

The paper is organized as follows: In Sect.\ 2 we give an overview of the AMBER observations of $\eta$ Car and 
describe the data reduction procedure in detail, and in Sect.\ 3, the analyses of the continuum data and the 
measurements within the Br$\gamma$ and \ion{He}{I} lines are discussed individually.

\section{AMBER observations and data processing}\label{sec:datared}

\begin{figure}[h]
\vspace*{-3mm}

\centering
  \includegraphics[width=62mm]{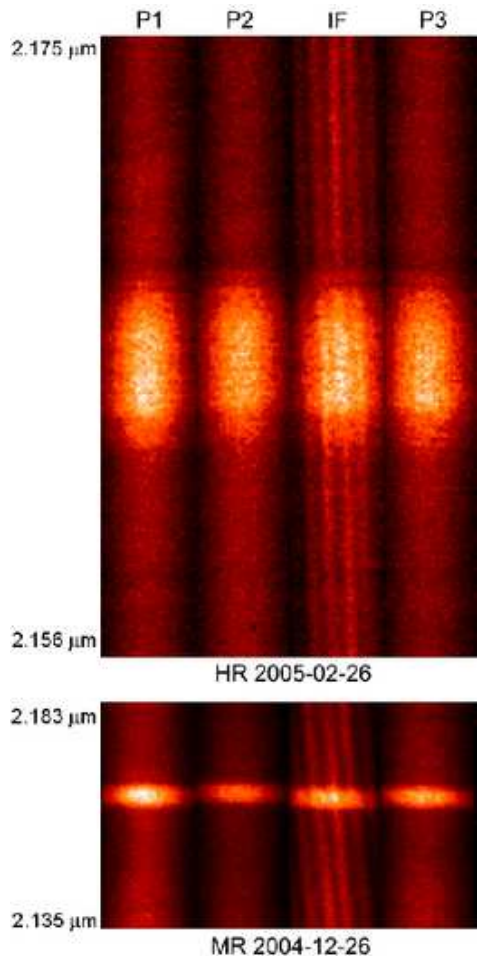}
  \caption{
Spectrally dispersed VLTI/AMBER Michelson interferograms of $\eta$ Car. The two panels show the spectrally 
dispersed fringe signal (IF) as well as the photometric calibration signals from the three telescopes (P1-P3) 
in high (HR, upper panel) and medium spectral resolution mode (MR, lower panel). In both panels, the bright 
regions are associated with the Doppler-broadened Br$\gamma$ emission line.
}
  \label{fig:rawdata}
\end{figure}

AMBER \citep{petrov03, pet06, petrov06} is the near-infrared ($J$, $H$, $K$ band) beam-combiner instrument
of ESO's \textit{Very Large Telescope Interferometer}, which allows the measurement of visibilities, 
differential visibilities, differential phases, and closure phases \citep{petrov03,millour06}.
AMBER offers three spectroscopic modes: low (LR mode; \textit{R}=$\lambda/\Delta\lambda$=75), medium 
(MR mode; \textit{R}=1,500), and high (HR mode; \textit{R}=12,000) spectral resolutions. The fibers in 
AMBER limit the field-of-view to the diameter of the fibers on the sky ($\sim60\,$mas). In AMBER the light 
is spectrally dispersed using a prism or grating. The AMBER detector is a Hawaii array detector with 
512$\times$512 pixels.

Figure~\ref{fig:rawdata} shows two AMBER raw interferograms taken in the wavelength range around the Br$\gamma$ 
line in HR (top) and MR (bottom) mode. In the MR data sets, the Doppler-broadened Br$\gamma$ line covers
$\sim 8$ spectral channels, whereas in HR mode, the line is resolved by $\sim 50$ spectral channels.

$\eta$~Car was observed with AMBER on 2004 December~26, 2005 February~25, and 2005 February~26 with the three 
8.2\,m \textit{Unit Telescopes} UT2, UT3, and UT4. With projected baseline lengths up to 89~m, an angular 
resolution of $\sim$5 mas was achieved in the $K$ band. As listed in Table~\ref{tab:observations}, the MR-K 
observations were performed in the wavelength range around both the \ion{He}{I}~2.059\,$\mu$m and the 
Br$\gamma$~2.166\,$\mu$m emission lines. The HR-K observations were only performed in a wavelength range around
the Br$\gamma$ line. The widths of the wavelength windows of the obtained MR-K and HR-K observations are 
approximately 0.05\,$\mu$m and 0.02\,$\mu$m, respectively.

\begin{figure*}[hp]
  \centering      
  \includegraphics[height=200mm]{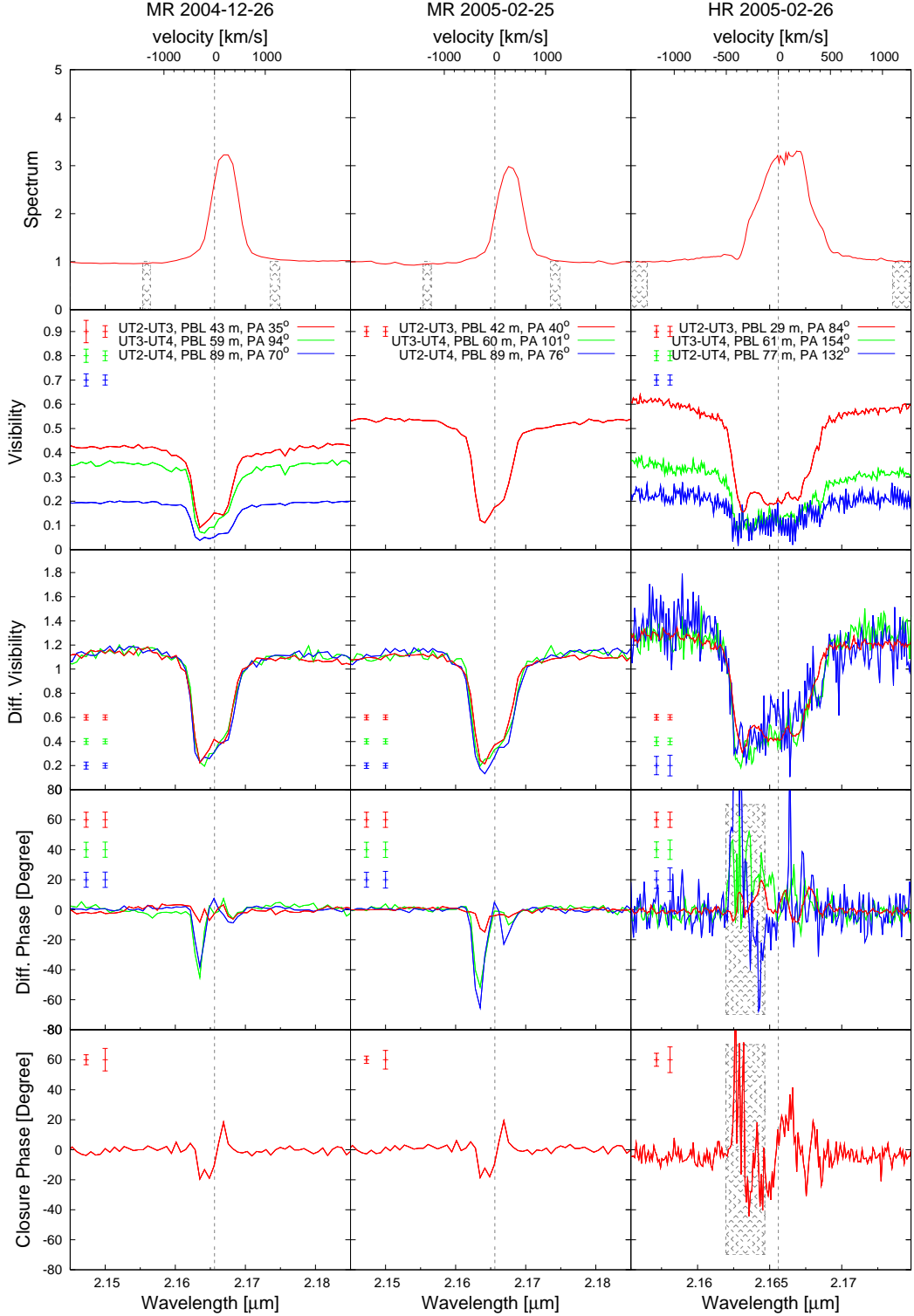}
  \caption{
    AMBER observables derived from our $\eta$~Car data around the Br$\gamma$ line for three independent 
    measurements (\textbf{Left:}~MR, 2004~December~26, \textbf{Middle:}~MR, 2005~February~25,
    \textbf{Right:}~HR, 2005~February~26). The first row shows the continuum-normalized spectra as extracted 
    from the interferometric channels, followed by the derived calibrated visibilities and the differential 
    visibilities. In the fourth and fifth row, the differential phase and the closure phase are presented.
    In the spectra we mark the wavelength regimes, which we defined as continuum for our analysis (shaded 
    regions). The vertical grey line marks the rest-wavelength of Br$\gamma$ ($\lambda_{\rm vac}=2.1661\,\mu$m; 
    the small correction due to the system velocity of -8 km/s \citep{smi04} has been neglected). We show 
    different error bars within each panel:  The left error bars correspond to the total (including statistical 
    and systematic) error estimated for the continuum wavelength range, and the error bar towards the right 
    visualizes the total error for the wavelength range within the line. For the HR 2005-02-26 measurement, 
    data splitting showed that for a small wavelength range (hatched areas in the two lower right panels), 
    the differential phase for the longest and middle baseline as well as the closure phase become very noisy 
    and are therefore not reliable. Furthermore, the HR differential phase of the longest baseline is noisy 
    at all wavelengths. See Sect.~\ref{sec:datared} for further details.
  }
  \label{fig:overview.BrG.ALL}
\end{figure*}

For the reduction of the AMBER data, we used version 2.4 of the \textit{amdlib}\footnote{This software package is 
available from\\ {\tt http://amber.obs.ujf-grenoble.fr}} software package. This software uses the P2VM (\textit{
pixel-to-visibilities matrix}) algorithm \citep{tatulli06} in order to extract complex visibilities for each baseline
and each spectral channel of an AMBER interferogram.  From these three complex visibilities, the amplitude and the 
closure phase are derived. While the closure phase is self-calibrating, the visibilities have to be corrected for
atmospheric and instrumental effects. This is done by dividing the $\eta$~Car visibility through the visibility of 
a calibrator star measured on the same night. In order to take the finite size of the calibrator star into account, 
the calibrator visibility is corrected beforehand through division by the expected calibrator star visibility (see
Table~\ref{tab:observations}).  In the case of the MR measurement performed on 2005-02-25, the interferograms 
recorded on the calibrator contain only fringes corresponding to the shortest baseline (UT2-UT3).  Thus, the
$\eta$~Car visibility for this night could only be calibrated for this shortest baseline. 

Besides the calibrated visibility and the closure phase, the spectral dispersion of AMBER also allows us to compute 
differential observables; namely the differential visibility and the differential phase \citep{petrov03,pet06,
petrov06,millour06}. These quantities are particularly valuable, as they provide a measure of the spatial extent 
and spatial offset of the line-emitting region with respect to the continuum emission. Since the measured complex 
visibilities are affected by wavelength-dependent atmospheric piston (optical path difference), the piston has to 
be estimated and subtracted. This was done using the \textit{ammyorick$^1$} tool (version 0.56).

Since a large fraction of the interferograms is of low contrast (probably due to vibration; see \citealt{malbet06}),
we removed a measurement from the data sets if \textit{(a)} the intensity ratio of two of the photometric channel 
signals is larger than 4 (a large ratio means that the interferograms are very noisy since the signal is very weak 
in one channel) or \textit{(b)} it belongs to the 70 percent of the interferograms with the lowest fringe contrast 
SNR (with the SNR defined as in \citealt{tatulli06}). In order to optimize the selection for each baseline of the 
telescope triplet, both of these criteria are applied for each telescope pair individually. Furthermore, the first 
10 frames in each new sequence of recorded interferograms are removed since they are degraded by electronic noise. 

\begin{figure}[h]
\hspace*{8mm}
  \includegraphics[width=70mm,angle=270]{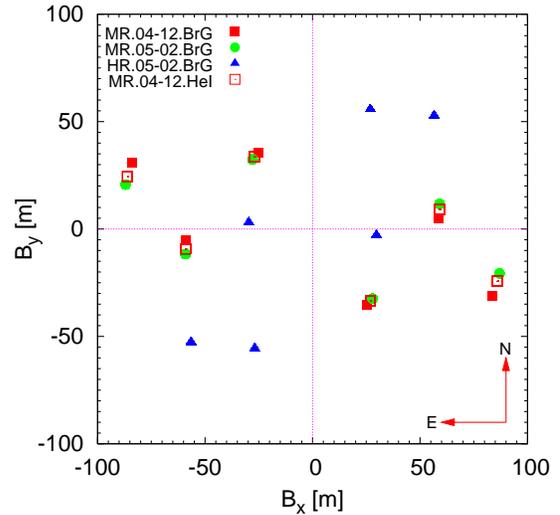}
  \caption{
uv coverage of the AMBER $\eta$~Car observations. The data obtained with medium spectral 
resolution in Dec.\ 2004 and Feb.\ 2005 are indicated by red squares and green bullets, 
respectively, while the high-resolution measurements are shown as blue triangles. Filled 
symbols denote observations around the Br$\gamma$ line, and open symbols denote observations 
around the \ion{He}{I} line.
}
  \label{fig:uvcov}
\end{figure}

Figures~\ref{fig:overview.BrG.ALL} and \ref{fig:overview.HeI.ALL} show the spectra as well as the wavelength 
dependence of the visibilities, differential visibilities, differential phases, and closure phases derived from 
the AMBER interferograms for the observations around the Br$\gamma$ and \ion{He}{I} emission lines. The uv coverage 
of the observations is displayed in Fig.~\ref{fig:uvcov}. 

The $\eta$~Car spectra were corrected for instrumental effects and atmospheric absorption through division by the 
calibrator spectrum. For the HR 2005-02-26 measurement, we found that the calibrator itself (L~Car) shows prominent 
Br$\gamma$ line absorption (see Fig.~\ref{fig:speccal}). Therefore, we had to remove this stellar line by linear 
interpolation before the spectrum could be used for the calibration. The wavelength calibration was done using 
atmospheric features, as described in more detail in Appendix~\ref{app:wavecal}.

In order to test the reliability of our results, we split each of the raw data sets into 5 subsets, each containing 
the same number of interferograms. The results obtained with these individual subsets allowed us to test that the 
major features detected in the visibility, differential visibility, differential phase, and closure phase are stable, 
even without any frame selection applied. As an exception, we found that for a small wavelength range of the HR 
2005-02-26 data set (hatched areas in the two lower right panels of Figure~\ref{fig:overview.BrG.ALL}), the 
differential phase corresponding to the middle and longest baselines and the closure phase vary strongly within the 
subsets and are, therefore, unreliable.  This is likely due to the very low visibility value on these two baselines, 
resulting in a low fringe SNR within this wavelength range. Furthermore, with this method we found that the 
differential visibility, differential phase, and closure phase extracted from the MR 2005-02-25 \ion{He}{I} data set 
are very noisy and not reliable. Therefore, these differential quantities and closure phases were dropped from our 
further analysis.

The subsets were also used to compute statistical errors. We estimated the variance for each spectral channel and 
derived formal statistical errors for both the continuum and line wavelength ranges. In each panel of 
Figures~\ref{fig:overview.BrG.ALL} and \ref{fig:overview.HeI.ALL}, we show two types of error bars corresponding 
to these regions, which not only take these statistical errors but also a systematic error (e.g.\ resulting from an
imperfect calibration) into account.  


\section{Observational results and interpretation}\label{sect:res}

\subsection{Comparison of the observed wave\-length de\-pen\-dence of the
  visibility with the NLTE radiative transfer model of Hillier et al. (2001)}\label{res:overview}

\begin{figure}[hp]
  \centering
  \includegraphics[height=220mm]{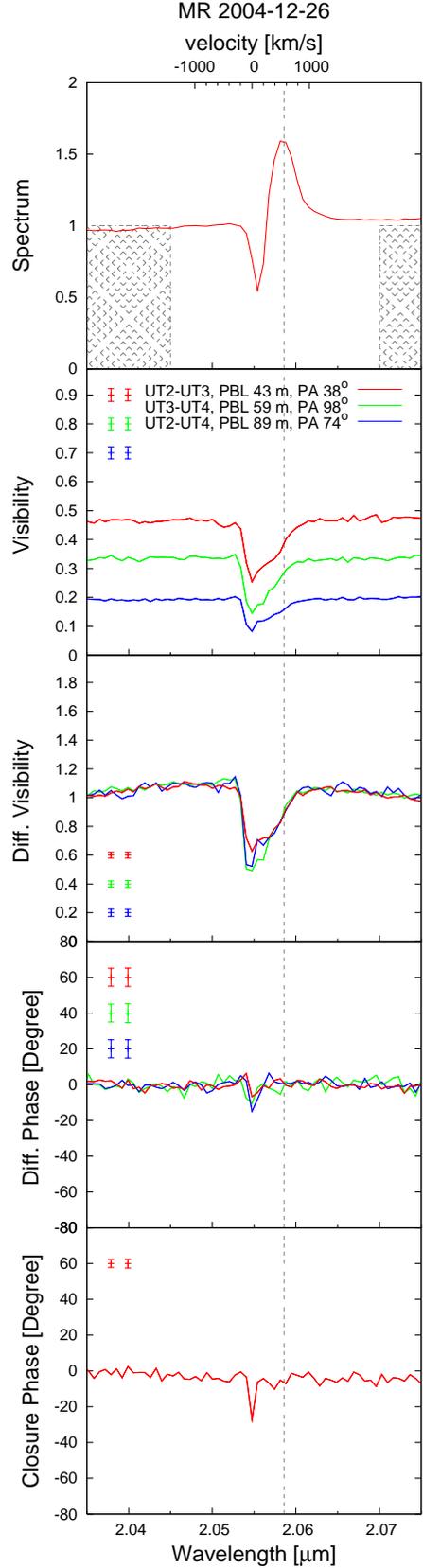}
  \caption{
    Similar to Fig.~\ref{fig:overview.BrG.ALL}, but showing the MR measurement from 2004~December~26 covering 
    the region around the \ion{He}{I} line. The vertical grey line marks the \ion{He}{I} rest-wavelength 
    ($\lambda_{\rm vac}=2.0586\,\mu$m).
  }
  \label{fig:overview.HeI.ALL}
\end{figure}

\begin{figure*}[t]

\vspace*{10mm}

\hspace*{10mm}
\includegraphics[width=115mm,angle=-90]{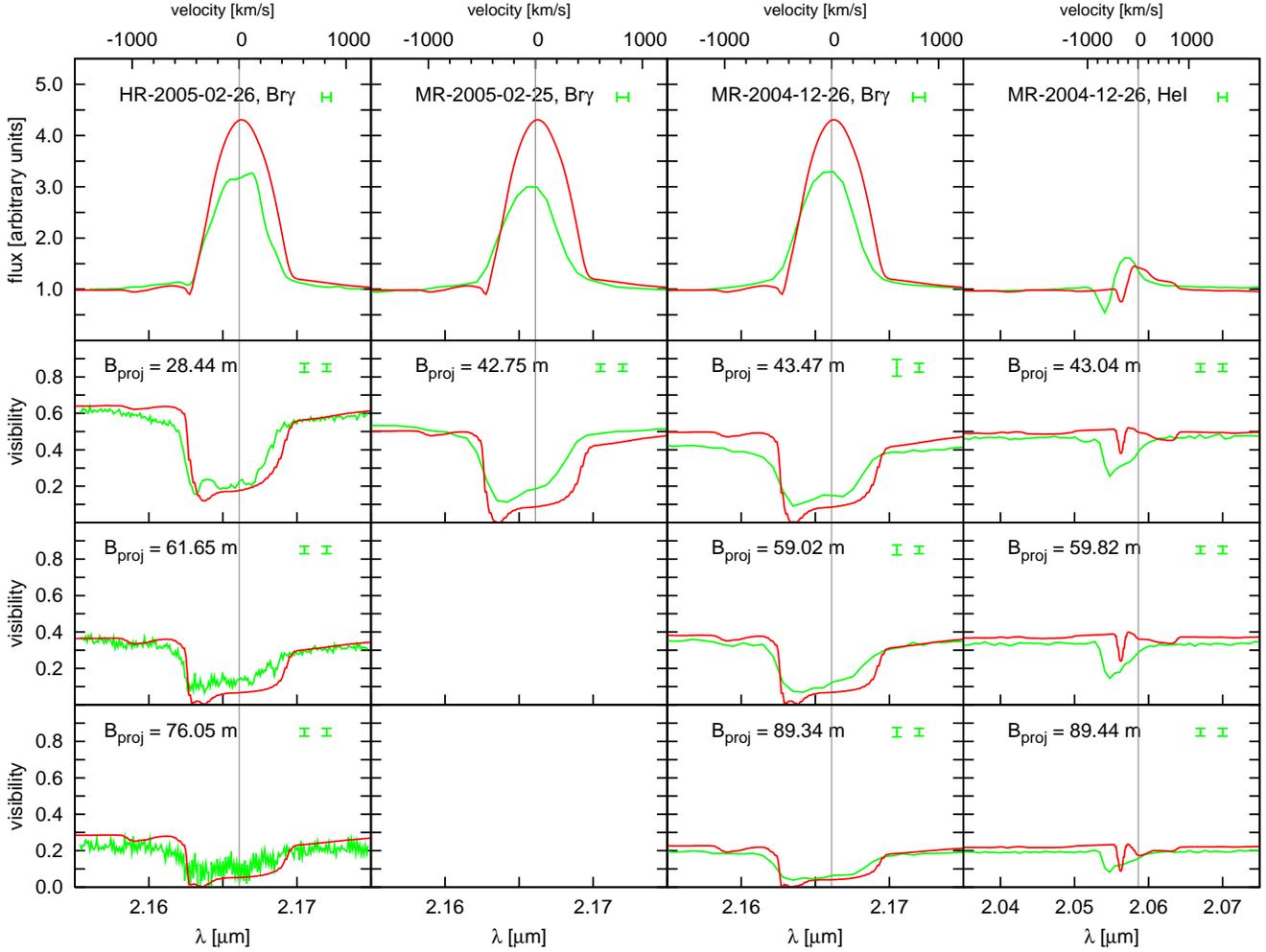}
\vspace*{8mm}

   \caption{\label{res_nlte_vislam}
Comparison of the AMBER spectra and visibilities with the NLTE model predictions of \citet{hillier01}. 
The figure displays the spectra (upper row) and visibilities (lower three rows, see labels for
projected baselines) of the four AMBER measurements (green lines) and the corresponding  data of 
the Hillier et al.\ NLTE model (red lines). The errors of the AMBER continuum and line visibility
measurements are indicated by the two vertical error bars (see Figs.~\ref{fig:overview.BrG.ALL} and 
\ref{fig:overview.HeI.ALL}; the left bar is the continuum error bar), and the uncertainty of the AMBER 
wavelength calibration is indicated by the horizontal error bar. As the figure shows, we find good 
agreement between the AMBER data and the model predictions for the continuum visibilities as well as 
the shape and depth of the visibility inside the Br$\gamma$ line. In the case of the \ion{He}{I} line, 
the wavelength dependence of the model visibility inside the line differs considerably from the AMBER 
measurements, indicating a {\it different physical process involved in the line formation} (see 
Sect.~\ref{res:overview}). The \ion{He}{I} wavelength shift can be attributed to a combination of both 
Doppler shift \citep[e.g.][]{nielsen06,hillier06} and uncertainties in the wavelength calibration of 
the AMBER data. Note that no additional scaling has been applied to the Hillier et al.\ model. The 
model spectra and visibilities have a spectral resolution ($\lambda/\Delta\lambda\sim20,000$) 
comparable to the HR measurements.
     }
    \end{figure*}

For the analysis presented in this chapter, we used the AMBER data sets from 2004 Dec.\ 26 and 2005 
Feb.\ 25 and 26, presented in Figs.\ \ref{fig:overview.BrG.ALL} and \ref{fig:overview.HeI.ALL}, and 
compared the AMBER visibilities and spectra with the NLTE radiative transfer model of \citet{hillier01}.
To directly compare the AMBER measurements with this model, we derived monochromatic model visibilities 
for all wavelengths between 2.03 and 2.18$\,\mu$m (with $\Delta \lambda =10^{-4}\,\mu$m) from the model 
intensity profiles, assuming a distance of 2.3 kpc for $\eta$~Car. The comparison is visualized in 
Fig.~\ref{res_nlte_vislam} for the individual AMBER HR and MR measurements. The first row displays the 
AMBER and model spectra, while all other panels show the AMBER and model visibilities for the different 
projected baselines. We note that for the comparison shown in Fig.~\ref{res_nlte_vislam}, we used the 
original model of \citet{hillier01} without any additional size scaling or addition of a background 
component.

As the figure reveals, the NLTE model of \citet{hillier01} can approximately reproduce the AMBER 
\textit{continuum} observations for all wavelengths (i.e.\ 2.03--2.18$\,\mu$m) and all baselines. 
Moreover, the wavelength dependence of the model visibilities inside the Br$\gamma$ line is also 
similar to the AMBER data. There is a slight tendency for the model visibilities in the Br$\gamma$ 
line to be systematically lower, which can be attributed to the overestimated model flux in the line. 
On the other hand, there is an obvious difference in the wavelength dependence of the visibility across 
the \ion{He}{I} line between the observations and the model predictions. This difference probably indicates 
that the primary wind model does not completely describe the physical origin and, hence, the spatial 
scale of the \ion{He}{I} line-forming region. The discrepancy is possibly caused by additional \ion{He}{I} 
emission from the {\it wind-wind interaction zone} between the binary components and by the {\it primary's 
ionized wind zone} caused by the secondary's UV light illuminating the primary's wind 
\citep[e.g.,][]{davidsonetal99,davidson01,pico02,sd04,hillier06,nielsen06,martin06}, as discussed in 
Sects.~\ref{sect_intro}, \ref{sect:res_hei}, and \ref{amber_performance} in more detail.

\begin{figure*}[ht]
\sidecaption
\includegraphics[width=120mm,angle=0]{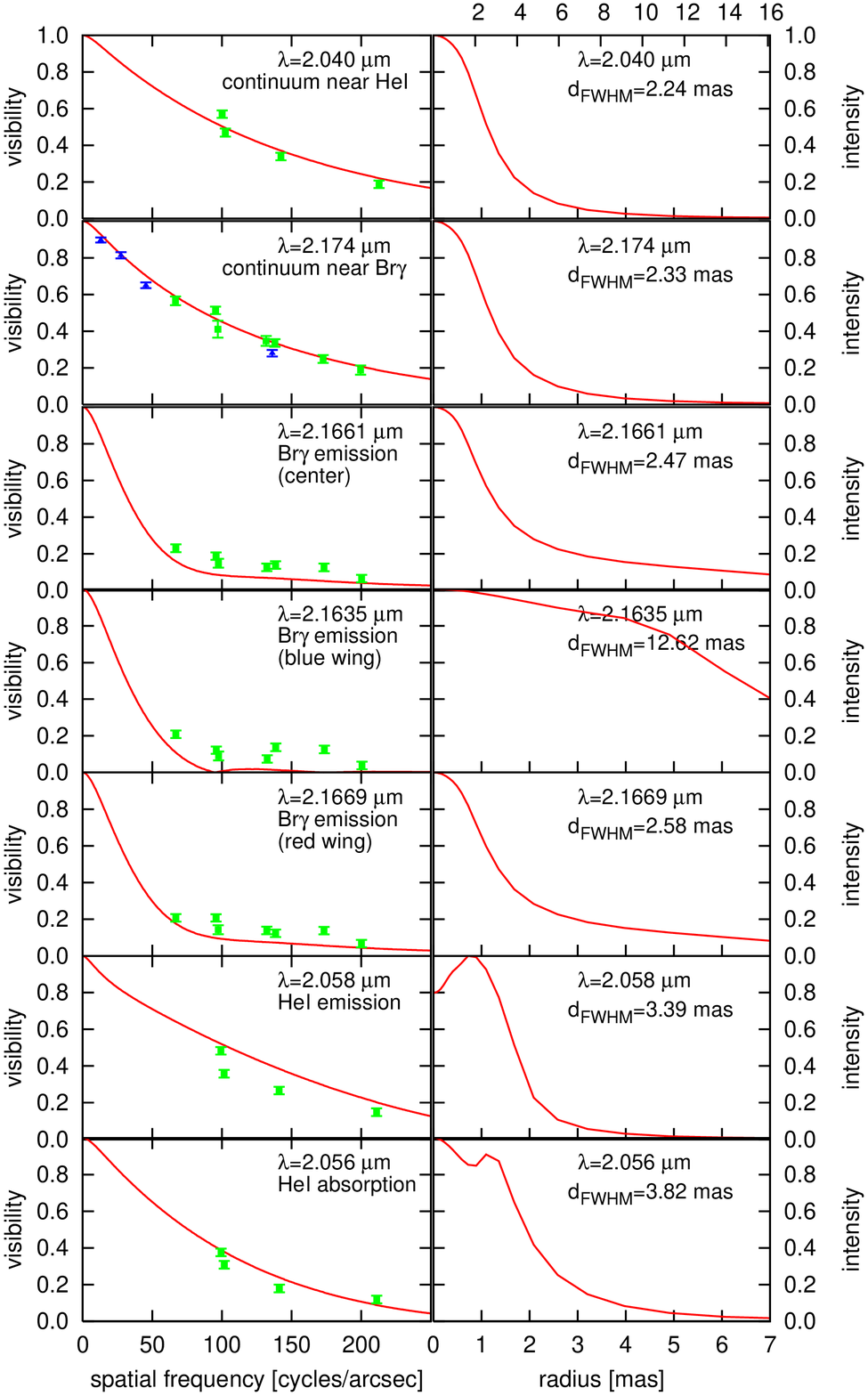}
%
\caption{\label{res_nlte_visclv}
{\bf Left}: 
Comparison of the AMBER visibilities (filled green squares; baseline range 28--89~m) as a function of
spatial frequency with the NLTE model predictions of \citet{hillier01} (solid red lines) for two continuum 
wavelengths (upper two panels; see labels for the exact wavelengths), the central wavelength of the 
Br$\gamma$ emission line (third), wavelengths in the blue and red wing of the Br$\gamma$ emission line 
(fourth and fifth row; at these wavelengths, the AMBER data show the strongest differential and closure 
phase signals), and the central wavelengths of the \ion{He}{I} emission and P Cygni absorption (lower 
two panels). The blue triangles are the background-corrected VINCI $K$-band measurements from \citet{van03}.
{\bf Right}:
Center-to-limb variation (CLV; i.e.\ intensity as a function of angular radius) of the monochromatic
\citet{hillier01} NLTE models for the wavelengths indicated by the labels. As the figures show, the
typical FWHM diameter of the models is of the order of $\sim$2--4~mas for an assumed distance of 2.3~kpc
for $\eta$~Car. See text for details.
}
\end{figure*}

Figure~\ref{res_nlte_visclv} shows the AMBER and model visibilities as a function of spatial frequency 
and the corresponding model center-to-limb intensity variations (CLVs) for seven selected wavelengths 
(2 continuum wavelengths; center, blue-shifted, and red-shifted wings of Br$\gamma$ emission; center of 
both \ion{He}{I} emission and absorption). As Fig.~\ref{res_nlte_visclv} reveals, at several wavelengths 
we find a very good agreement between the visibilities measured with AMBER and the visibilities predicted 
by the model of \citet{hillier01}. This is especially true for the continuum data (upper two panels). 

>From the model CLVs, FWHM model continuum diameters of 2.24~mas and 2.33~mas can be derived for 
$\lambda=2.040$ and 2.174\,$\mu$m, respectively. If we allow for a moderate rescaling of the size of 
the model, we find that the best $\chi^2$ fit at both continuum wavelengths can be obtained with 
scaling factors of 1.015 and 1.00, respectively. This means that the model size has to be increased 
by only 1.5\% at $\lambda=2.040\,\mu$m and that the best fit at 2.174\,$\mu$m is indeed obtained with 
the original Hillier model with a scaling factor of 1.0. Thus, taking the slight rescaling for the
best $\chi^2$ fit into account, we can conclude that, based on the NLTE model from \citet{hillier01,hillier06}
and the AMBER measurements, the apparent FWHM diameters of $\eta$ Car in the $K$-band continuum at 
$\lambda=2.040\,\mu$m and 2.174\,$\mu$m are 2.27~mas and 2.33~mas, respectively (see Table 2),
corresponding to a physical size of approximately 5~AU.

Since the deviations between the model and the measurements are larger in the case of the Br$\gamma$
and \ion{He}{I} line data (lower 5 panels in Fig.~\ref{res_nlte_visclv}), the scaling factors corresponding
to the best $\chi^2$ fit in the lines show stronger deviations from unity. For the Br$\gamma$ emission 
line, we find scaling factors of 0.74, 0.76, and 0.78 for $\lambda = 2.1661, 2.1635,$ and $2.1669\,\mu$m, 
corresponding to FWHM diameters of 1.83, 9.52 and 2.02 mas (see Table 2). 

For the \ion{He}{I} emission line, rescaled models with scaling factors of 1.24 and 1.11 provide the best 
$\chi^2$ fit for the peaks of the emission and absorption within the \ion{He}{I} line ($\lambda=2.058$ 
and $2.056\,\mu$m), resulting in FWHM diameters of 4.24 and 4.19~mas, respectively.

In addition to the inner CLV core, at several wavelengths, the CLVs show a very extended wing corresponding 
to the extended Br$\gamma$ and \ion{He}{I} line emission regions. Since the intensity in the wing is much 
lower than 50\% of the peak intensity, the FWHM diameter is not very sensitive to this part of the CLV.  
In other words, in the case of CLVs with multiple or very extended components, a FWHM diameter can be quite 
misleading. In such a case, it seems to be more appropriate to use, for instance, the diameter measured at 
10\% of the peak intensity ($d_{10\%}$) or the {\it 50\% encircled-energy diameter} ($d_{\rm 50\%\,EED}$).
For example, at $\lambda=2.1661\,\mu$m we obtain $d_{10\%}=9.39$~mas and $d_{\rm 50\%\,EED}=9.58$~mas, 
while for the continuum at $2.174\,\mu$m we find $d_{10\%}=5.15$~mas and $d_{\rm 50\%\,EED}=4.23$~mas. 
Thus, based on $d_{\rm 50\%\,EED}$, $\eta$~Car appears $\sim2.2$ times larger at $\lambda=2.1661\,\mu$m 
compared to the continuum at $\lambda=2.174\,\mu$m. The best-fit model diameters at the other wavelengths 
are listed in Table 2. The errors of the diameter measurements are $\pm4\%$ for the two continuum diameters 
and $\pm10\%$ for the line diameters, derived from the visibility errors and the uncertainty of the fitting 
procedure.

\begin{table}[t]
\caption{
Diameters obtained by fitting \citet{hillier01} model visibilities
to the measured AMBER visibilities. Errors are $\pm4\%$ for the diameters in the continuum
and $\pm10\%$ in the lines (see text).
}
\label{tab:diameters}
\begin{tabular}{llrrr}
\hline\hline
Spectral region   & Wavelength & $d_{\rm FWHM}$ & $d_{\rm 10\%}$ & $d_{\rm 50\%\,EED}$ \\
                  & [$\mu$m]   & [mas]          & [mas]          & [mas]             \\
\noalign{\smallskip} \hline \noalign{\smallskip} 
continuum                 & 2.0400   & 2.27        & 4.85      & 3.74       \\
continuum                 & 2.1740   & 2.33        & 5.15      & 4.23       \\
Br$\gamma$ (center)       & 2.1661   & 1.83        & 9.39      & 9.58       \\
Br$\gamma$ (blue wing)    & 2.1635   & 9.52        & 16.46     & 9.60       \\
Br$\gamma$ (red wing)     & 2.1669   & 2.02        & 9.61      & 9.78       \\
\ion{He}{I} (absorption)  & 2.0560   & 4.24        & 8.22      & 5.36       \\
\ion{He}{I} (emission)    & 2.0580   & 4.19        & 4.30      & 6.53       \\        
\noalign{\smallskip} \hline
\end{tabular}\newline
$d_{\rm FWHM}$ = FWHM diameter;
$d_{\rm 10\%}$ = diameter measured at 10\% peak intensity;
$d_{\rm 50\%\,EED}$ = 50\% encircled-energy diameter.

\end{table}

\subsection{Continuum visibilities}\label{sect:res_cont}
\subsubsection{Comparison of the continuum visibilities with the Hillier et al.\ (2001) 
               model predictions}\label{hilcontcomp}

The comparison of the AMBER {\it continuum} visibilities with the NLTE model from 
\citet{hillier01,hillier06} is shown in the two upper left panels of Fig.~\ref{res_nlte_visclv} 
for the continuum near the \ion{He}{I} $2.059\,\mu$m and Br$\gamma$ 2.166$\,\mu$m emission lines 
(the exact wavelengths are described in Fig.~6). Taking a slight rescaling into account, we 
concluded in the previous section that, based on the NLTE model from \citet{hillier01} and the 
AMBER measurements, the apparent 50\% encircled-energy diameters $d_{\rm 50\%\,EED}$ of $\eta$ Car 
in the $K$-band continuum at $\lambda=2.040\,\mu$m and 2.174\,$\mu$m are 3.74~mas and 4.23~mas, 
respectively (see Table 2). These diameters are in good agreement with the 50\% encircled-energy 
$K$-band diameter of 5~mas reported by \citet{van03}.

For comparison, we also fitted the AMBER visibilities with simple analytical models such as Gaussian
profiles, as described in more detail in Sect.~\ref{app:udgauss} in the Appendix. From a Gaussian fit
of the AMBER visibilities, we obtain a FWHM diameter of $d_{\rm Gauss}\sim 4.0\pm0.2\,$mas in the 
$K$-band continuum. As outlined in Sect.~\ref{app:udgauss}, the diameter value strongly depends on 
the range of projected baselines used for the fit, since a Gaussian without an additional background 
component is not a good representation of the visibility measured with AMBER. As discussed in 
Appendix~\ref{app:udgauss}, using a Gaussian fit with a fully resolved background component as a 
free parameter results in a best fit with a 30\% background flux contribution \citep[see also][]{pet06}.
 
\subsubsection{Comparison of the VINCI and AMBER continuum visibilities}\label{sect:vinci_verg}

In Fig.~\ref{res_nlte_visclv} (left, second row) displaying the averaged Br$\gamma$ continuum data, 
the visibilities of $\eta$ Car obtained with VLTI/VINCI are shown in addition to the AMBER data. 
These VINCI measurements were carried out in 2002 and 2003 using the 35~cm test siderostats at the 
VLTI with baselines ranging from 8 to 62~m \citep[for details, see][]{van03}. Like AMBER, VINCI is a 
single-mode fiber instrument. Therefore, its field-of-view is approximately equal to the Airy disk of 
the telescope aperture on the sky, which is $\sim1.4^{''}$ in the case of the siderostats. From the 
VINCI measurements and using only the 24~m baseline data, \citet{van03} derived a FWHM Gaussian diameter
of $d_{\rm Gauss}\sim$~7 mas for the wind region of $\eta$~Car. At first glance, this diameter measurement 
seems to contradict the $d_{\rm Gauss}\sim 4.0$~mas FWHM diameter derived from the AMBER data. This is 
not the case, however, since the diameter fit is very sensitive to the baseline (or spatial frequency) 
fit range, because a Gaussian is not a good representation of the visibility curve at all, as can be 
seen in Fig.~\ref{res:allvud}. If only the VINCI data points are fitted, which have spatial frequencies 
$<$ 60 cycles/arcsec (corresponding to projected baselines $<$ 28~m), $d_{\rm Gauss}\sim 7$~mas provides 
the best fit. On the other hand, if the data point at 136 cycles/arcsec (corresponding to a projected 
baseline of $\sim62$~m) is included in the fit, we obtain $d_{\rm Gauss}\sim 4.3\,$mas (see also the 
discussion Sect.~\ref{app:udgauss}). Thus, when using comparable baseline ranges for the Gaussian fits, 
there is good agreement between the AMBER and VINCI measurements.

To account for the background contamination of the VINCI data caused by nebulosity within VINCI's large 
1.4$^{''}$ field-of-view (in which, for instance, all speckle objects B, C, and D are located), van 
Boekel et al.\ introduced a background component (derived from NACO data) providing 55\% of the total 
flux. Adding this background component to the model of \citet{hillier01}, they found a good match between 
the model and the observations. Since our AMBER observations were carried out with the 8.2~m Unit 
Telescopes of the VLTI, the field-of-view of the AMBER observations was only $\sim$60~mas. Thus, the 
background contamination of the AMBER data can be expected to be much weaker, if not negligible, compared 
to the VINCI measurements. To check this, we first performed a fit of the \citet{hillier01} model, which 
not only contains the size scaling as a free parameter, but also a fully resolved background component.
As we expected, we found the best fit (smallest $\chi^2$) with no background contamination.
\footnote{ An additional argument in favor of only a very faint background contribution in the AMBER UT 
observation can be found in the shape of the high spectral resolution line: the light from the speckle 
objects B, C, and D is produced in areas with velocities smaller than 50~km/s. Therefore, it produces 
a narrow emission line which should appear in the center of the broad Br$\gamma$ line. Just looking at 
the shape of the line, it can be concluded that such an effect is negligible.
}
Therefore, when we finally compared the AMBER observations with the model from \citet{hillier01}, we 
did not introduce a background component. In Fig.~\ref{res_nlte_visclv} (second row, left) we plot both 
the AMBER visibilities (no background correction required) plus the background-corrected VINCI data 
(assuming a 55\% background contribution; blue triangles). As can be seen from the figure, these VINCI 
points nicely match the AMBER data and the corresponding fit of the NLTE model from \citet{hillier01}. 
Therefore, from the analysis of the continuum data, we can conclude that the background contamination 
in the AMBER measurements is negligible and that the AMBER measurements are in good agreement with both 
the previous VINCI measurements and the model predictions from \citet{hillier01}. 

\subsection{Elongated shape of the continuum intensity distribution}\label{res_asymm}

To look for detectable elongations of the continuum intensity distribution, we fitted an elliptically 
stretched 2-D version of the radiative transfer model visibilities from \citet{hillier01} to the measured 
visibilities. Our best $\chi^2$ fit reveals a projected axis ratio of $\xi\,=\,1.18\pm0.10$ and PA 
$=\,120\pm15$\degr. Comparison with the results found by \citet{van03} shows that the projected axis 
ratio $\xi$ derived from the AMBER data is in basic agreement with the $K$-broad-band values of 
$\xi\,=\,1.25 \pm 0.05$ and PA$\,=\,138\pm7\degr$ from \citet{van03}.

We also studied the elongation inside the Br$\gamma$ emission line at $\lambda=2.166\,\mu$m, following 
the same procedure as in the continuum; i.e., we fitted an elliptically stretched 2-D version of the 
Hillier et al.\ model shown in Fig.~\ref{res_nlte_visclv} to the AMBER data. However, since the global 
shape of the model function at $\lambda=2.166\,\mu$m shows stronger deviations from the measurements 
than in the continuum, the elongation determination suffers from larger uncertainties, resulting in 
large error bars of the fit parameters. For instance, for $\lambda=2.166\,\mu$m we obtained 
$\xi=1.66\pm0.60$ and PA=81$\pm40$\degr\, from the best ellipse fit. 

The 2-D ellipse fitting was also performed for the continuum near the \ion{He}{I} emission line and in 
the center of the \ion{He}{I} line ($\lambda=2.057\,\mu$m), where our model fits give an axis ratio of 
$\xi = 1.35\pm0.30$ and a PA of the major axis of $98\pm40\degr$ in the continuum, and $\xi = 1.74\pm0.60$ 
and PA = $159\pm40\degr$ in the center of the \ion{He}{I} emission line. It should be noted that for 
the \ion{He}{I} line region, only four visibility points are available, covering the small PA range 
of only $60\degr$. Because of this limited number of data points and the small PA coverage, we conclude 
that the \ion{He}{I} elongation measurements in the continuum as well as the line region are not reliable 
and abandoned  in the further elongation analysis of the \ion{He}{I} data. 

>From the {\it K}-band VINCI data, \citet{van03} derived a PA of $138\pm7\degr$ for the major axis, very 
well aligned with the Homunculus \citep[$132\degr$,][]{davidsonetal01} and in agreement with our results 
(PA $= 120\pm15$\degr). Van  Boekel's and our continuum elongation measurements favor the physical model 
according to which $\eta$ Car exhibits an enhanced mass loss in polar direction as proposed, for instance, 
by \citet{owo96,owo98} or \citet{md01} for stars rotating close to their critical rotation speed. Axis 
ratios of the order of 1.2 appear reasonable in the context of such polar-wind models. Suppose, for example, 
that the wind's polar/equatorial density ratio is 2 at any given radius $r$, as reported by \cite{smi03} 
to explain latitude-dependent changes in the Balmer line profiles. Relevant absorption and scattering 
coefficients have radial dependencies between $n_e \sim r^{-2}$ (Thomson scattering) and $n_{e}^{2} 
\sim r^{-4}$ (most forms of thermal absorption and emission).  A meridional map of projected optical 
thickness through the wind would show cross-sections of prolate spheroids, correlated with the appearance 
of the configuration. With the radial dependencies and polar/equatorial density ratio mentioned above, 
these spheroids have axial ratios between about 1.2 and 1.4; i.e., appreciably less than 2. Viewed from an
inclination angle $i \approx 45$\degr \citep{davidsonetal01}, the apparent (projected) axis ratios are 
between 1.1 and 1.2.  This is merely one example, and we have omitted many details, but it illustrates that 
the polar/equatorial density ratio is around 2, in agreement with \citet{smi03}.

Finally, \citet{smi03} suggested that the stellar wind should become basically spherical during an event 
at periastron. This prediction can be tested if VLTI/AMBER data are obtained at the next periastron passage.

\begin{figure}[ht]
\centering
\includegraphics[width=90mm,angle=0]{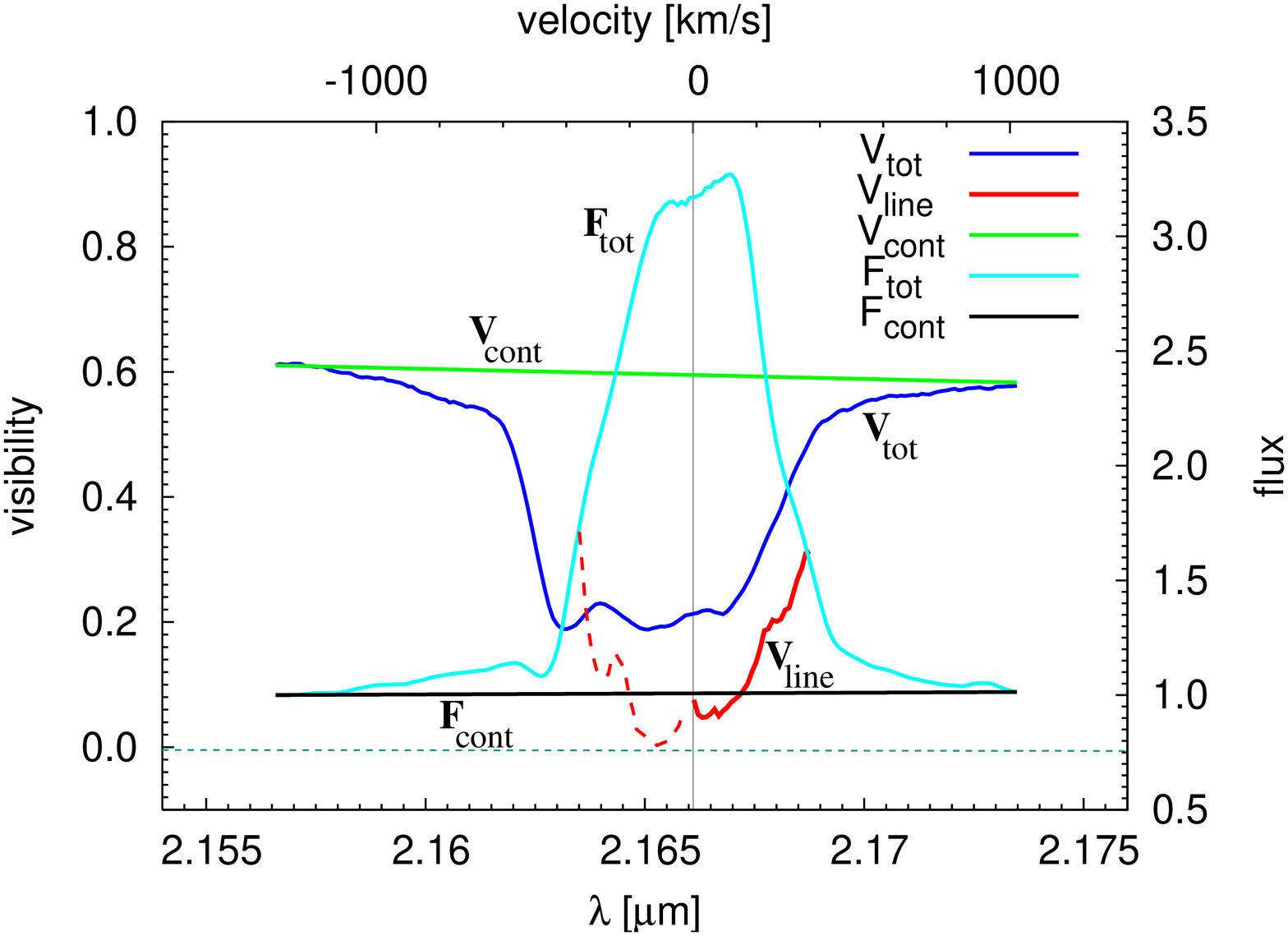}\\
\includegraphics[width=60mm,angle=-90]{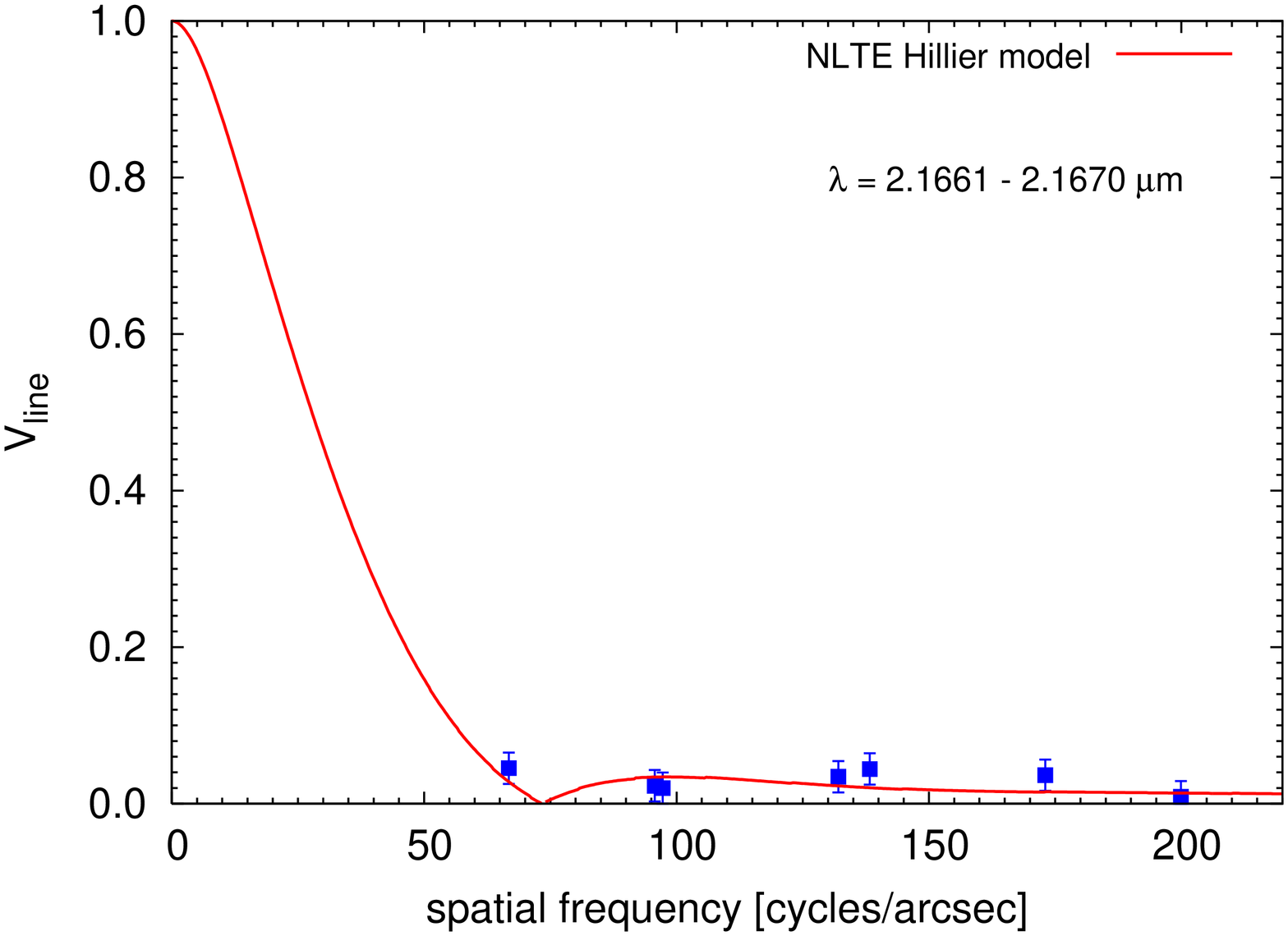}
   \caption{\label{res_nlte_brg}
{\bf Top}: 
The solid and dashed red lines show the continuum-corrected visibility inside the Br$\gamma$ line 
of the HR measurement with the shortest projected baseline (29~m). The continuum correction follows 
Eq.~(\ref{equ2}). Except for the phase, all quantities entering Eq.~(\ref{equ2}) are indicated 
in the figure. We assume $F_{\rm tot} = F_{\rm cont} + F_{\rm line}$ and that $F_{\rm cont}$ inside 
the line is approximately equal to the level outside the line. The continuum-corrected visibility 
in the blue-shifted region of the emission line is shown with a dashed line since it is highly 
uncertain due to the presence of the P Cygni-like absorption component (see text for details).
{\bf Bottom}:
Continuum-corrected AMBER visibilities (filled blue squares) in the red region of the Br$\gamma$ 
line as a function of spatial frequency. To derive the visibilities according to Eq.~(\ref{equ2}),
the data in the wavelength range 2.1661--2.1670$\,\mu$m (red-shifted line region) were averaged 
before the continuum correction. The solid red line shows the continuum-corrected model visibility. 
Obviously, observations at shorter baselines are needed to further test the model predictions.
}
    \end{figure}

\subsection{Continuum-corrected visibilities}\label{sect:contcorr}

\subsubsection{Continuum-corrected visibility in the Br\boldmath$\gamma\,$\boldmath emission line}
\label{sect:res_brg}


To investigate the brightness distribution in the Br$\gamma$ line in more detail, we tried to 
disentangle the continuum and pure line emission from both the AMBER data as well as the model 
data to derive the size of the pure Br$\gamma$ line-emitting region. Since the visibility measured 
inside an emission or absorption line is the composite of a pure line component and an underlying 
continuum, the measured line visibility, $V_{\rm tot}$ (see Fig.~\ref{res_nlte_brg} top), has to be 
corrected for the continuum contribution to obtain the visibility $V_{\rm line}$ of the line emitting 
(absorbing) region. As discussed in \citet{malbet06}, $V_{\rm line}$ can be calculated if the 
continuum level within the line is known. If the continuum within the line is equal to the continuum 
level outside the line for an optically thin environment, as assumed in Fig.~\ref{res_nlte_brg}, 
we obtain: 
\begin{equation}\label{equ1} 
V_{\rm line} = \frac{V_{\rm tot}\cdot F_{\rm tot} - V_{\rm cont}\cdot F_{\rm cont}}{F_{\rm line}}
\end{equation}

with $F_{\rm tot} = F_{\rm cont} + F_{\rm line}$ being the total measured flux and $V_{\rm tot}$
being the measured visibility (also see Fig.~\ref{res_nlte_brg} top for illustration). As outlined 
in the Appendix, taking a non-zero differential phase $\Phi^\prime$ into account introduces
an additional term in Eq.~(\ref{equ1}), leading to

\begin{equation}\label{equ2} 
V_{\rm line} = \frac{\sqrt{|V_{\rm tot}F_{\rm tot}|^2 + |V_{\rm cont}F_{\rm cont}|^2 
              - 2 V_{\rm tot}F_{\rm tot}V_{\rm cont}F_{\rm cont}\cdot \cos\Phi^\prime}}{F_{\rm line}}
\end{equation}

We applied Eq.~(\ref{equ2}) to the line-dominated AMBER data in the 2.155-2.175$\,\mu$m Br$\gamma$ 
wavelength range to derive the continuum-corrected visibility of the region emitting the Br$\gamma$ 
line radiation. One \textit{big uncertainty} in this correction is the unknown continuum flux within 
the line. Due to intrinsic absorption, the continuum flux might be considerably lower within the line 
than measured at wavelengths outside the line. Especially the blue-shifted wing of the Br$\gamma$ 
emission line might be affected by the P Cygni-like absorption, as discussed below. In case such an 
absorption component is present, our continuum-correction would overestimate the size of the 
line-emitting region. The influence of this effect on the red-shifted wing of the line is likely to 
be much smaller, as P Cygni-like absorption mainly affects the blue-shifted emission. 

The visibility across the Br$\gamma$ line is shown in Fig.~\ref{res_nlte_brg} (top) for the HR data 
corresponding to the shortest projected baseline. The results for the other data sets are similar.
>From Fig.~\ref{res_nlte_brg} (top), one can see that after the subtraction of a continuum contribution 
equal to the continuum outside the line, the visibility reaches very small values in the center of the 
emission line. This means that the pure line-emitting region is much larger than the region providing 
the continuum flux.

Fig.~\ref{res_nlte_brg} (top) shows a strong {\it asymmetry} between the blue- and red-shifted part of 
the visibility in the line with respect to the spectrum. While the visibility ($V_{\rm tot}$ as well as 
$V_{\rm line}$) rises concomitantly with the drop of the line flux on the red side, the situation is 
very different on the blue side line center. In agreement with the model predictions from \citet{hillier01}, 
this indicates the existence of a P Cygni-like absorption component in this wavelength region. In fact, 
at $\lambda=2.1625\,\mu$m, we see a small dip in the Br$\gamma$ spectrum in both the model spectrum from 
\citet{hillier01} and the HR AMBER observations. If such an absorption component is present, it can 
explain the asymmetric behaviour of the line visibility with respect to the spectrum. The P Cygni 
absorption in the blue wing of the Br$\gamma$ line makes the continuum correction of the visibility 
uncertain for wavelengths shorter than the central wavelength $\lambda_{\rm c}$ of the emission line.
Because of this uncertainty, in Fig.~\ref{res_nlte_brg} (top) the continuum corrected visibility is shown 
with a dashed line for $\lambda < \lambda_{\rm c}$, and the following discussion is restricted to the 
red-shifted region of the Br$\gamma$ line emission.

The continuum-corrected AMBER visibilities in the red-shifted region of the Br$\gamma$ line are displayed 
in Fig.~\ref{res_nlte_brg} (bottom) for all data sets. To derive the visibilities in the red region, the 
data in the wavelength range 2.1661--2.1670$\,\mu$m were averaged before the continuum correction. To now 
compare the continuum-corrected AMBER visibilities with the model predictions (2.1661--$2.1670\,\mu$m), 
we constructed a model intensity profile of the pure Br$\gamma$ emission line region by subtracting 
the Hillier et al.\  intensity profile of the nearby continuum from the combined line + continuum profile. 

As Fig.~\ref{res_nlte_brg} illustrates, the model prediction is in agreement with the low visibilities 
found for spatial frequencies beyond 60 cycles/arcsec. On the other hand, the figure also clearly indicates 
that measurements at smaller projected baselines are needed to further constrain the Hillier et al.\ model 
in the line-emitting region. With the baseline coverage provided by the current AMBER measurements, we 
obtain a FWHM diameter of $\gtrsim 15.4$~mas (lower limit) for the (continuum-corrected) line-emitting
region in the red line wing.

\subsubsection{Continuum-corrected visibility in the \ion{He}{I} emission line}\label{sect:res_hei}

As can be seen in Fig.~\ref{fig:overview.HeI.ALL}, the AMBER spectrum of the \ion{He}{I} line shows a 
P Cygni-like profile with a prominent absorption and emission component. This is in agreement with earlier 
findings by \citet{smi02} from long-slit spectroscopy using OSIRIS on the CTIO 4m telescope.

\begin{figure}[t]
\centering
\includegraphics[width=90mm,angle=0]{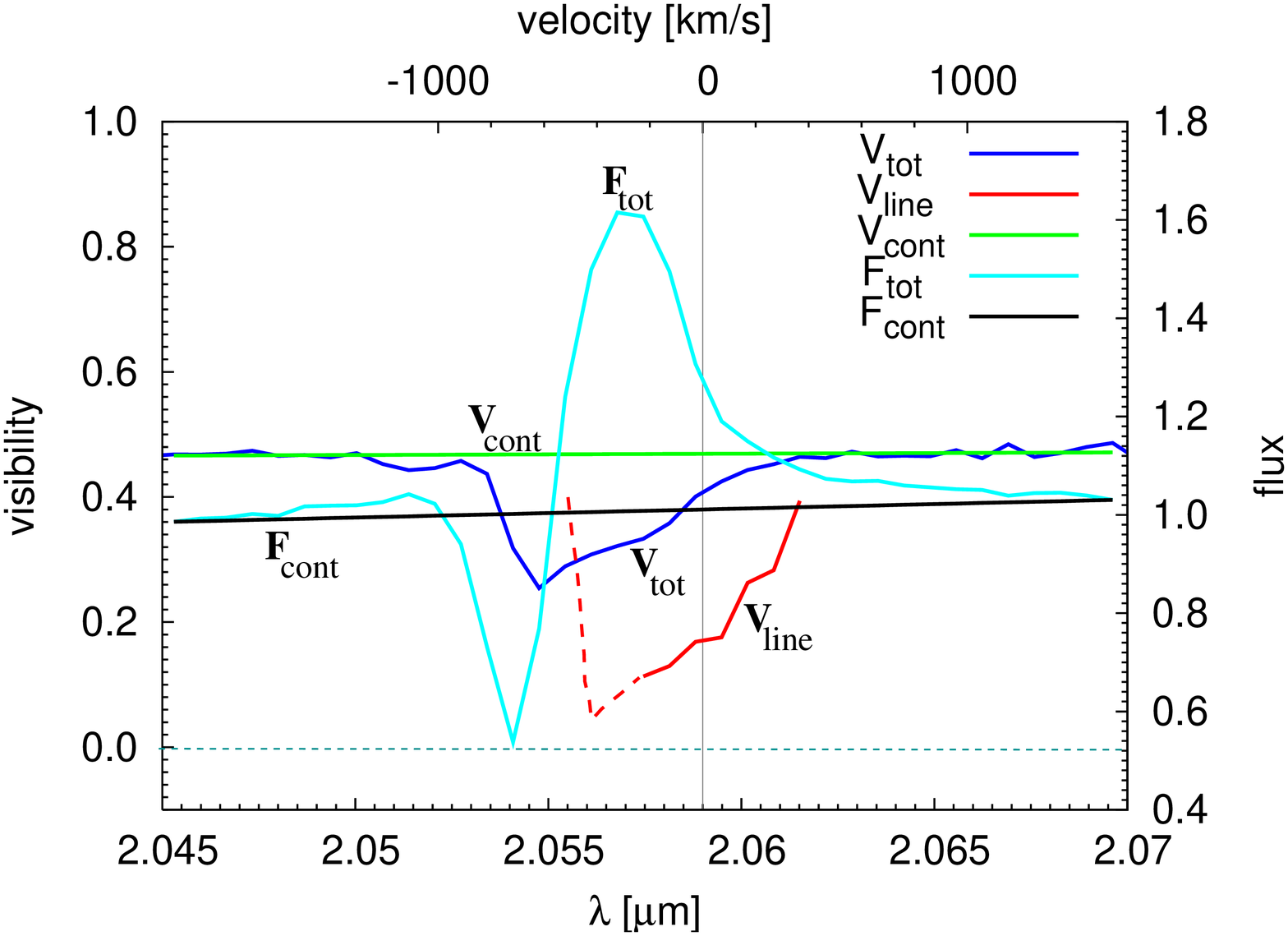}\\
\includegraphics[width=60mm,angle=-90]{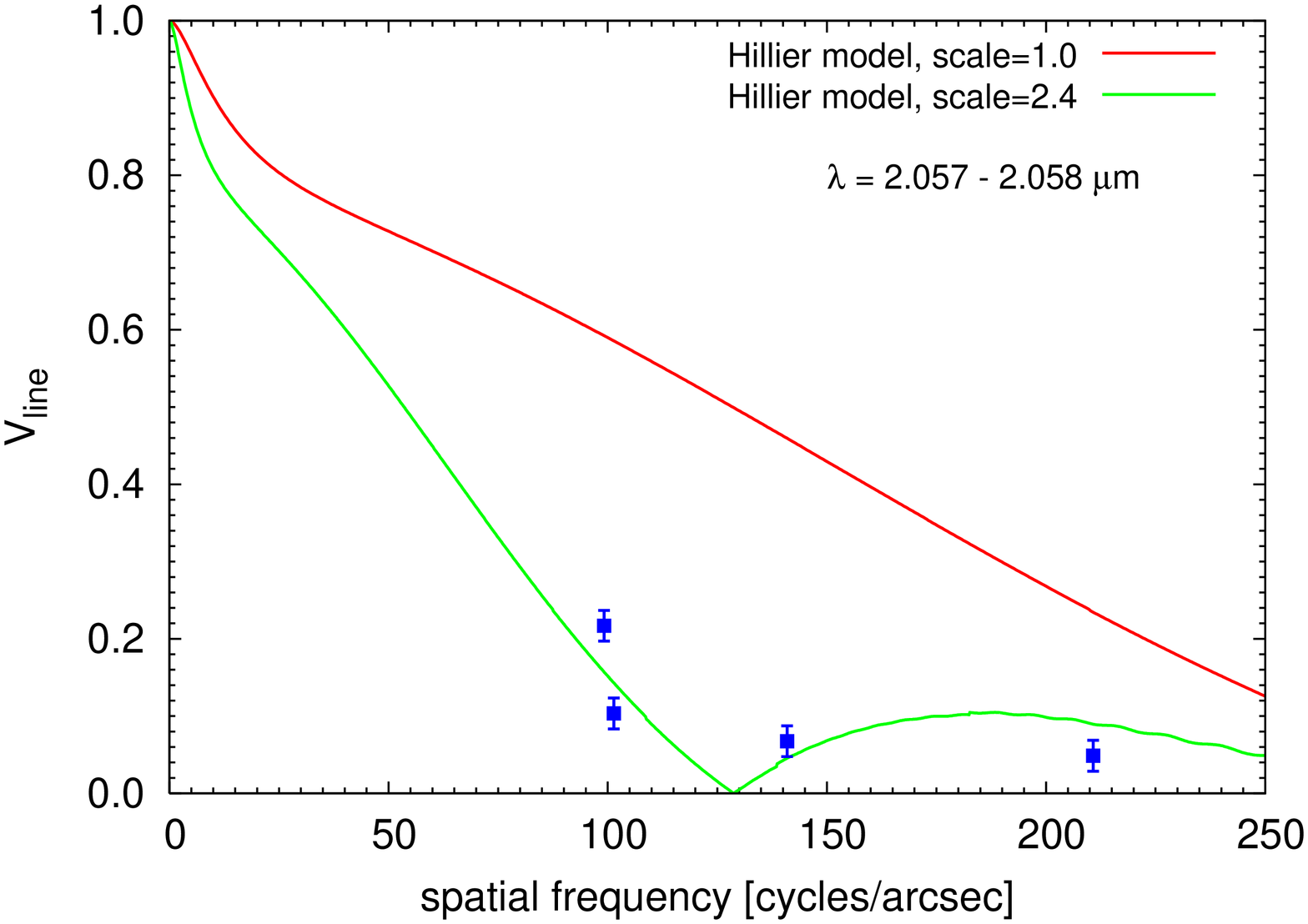}
   \caption{\label{res_heicontsubline}
Visibility in the \ion{He}{I} line. 
{\bf Top}:
The figure is similar to Fig.~\ref{res_nlte_brg} (top) but displays the MR-2004-12-26 data with the 
shortest projected baseline (43m). As in the case of the Br$\gamma$ emission, the figure reveals that 
the region dominated by line emission is fully resolved by the AMBER measurements. As in 
Fig.~\ref{res_nlte_brg} (top), the continuum-corrected visibility is shown with a dashed line in the 
blue-shifted region of the emission line to indicate that in this region, the line visibility is highly
uncertain due to the presence of the P Cygni absorption.
{\bf Bottom}: 
Continuum-corrected AMBER visibilities (filled blue squares) in the red region of the \ion{He}{I} 
emission line as a function of spatial frequency and continuum-corrected visibility curve (solid red line) 
according to the NLTE model of \citet{hillier01}. To derive the visibilities according to Eq.~(\ref{equ2}),
the data in the wavelength range 2.057--2.058$\,\mu$m were averaged before the continuum correction. The 
model curve predicts much larger visibility values in the line, which indicates that the size of the 
line-emitting region in the model is probably underestimated by the model. A rescaling of the size of the 
model by a factor of $\sim$2.3 (green line) reveals better agreement with the AMBER measurements. 
}
              \label{res_vishei}
\end{figure}

To estimate the spatial scale of the region emitting the \ion{He}{I} emission line, we followed the 
same approach as outlined in the previous section for the Br$\gamma$ line; i.e., we first applied 
Eq.~(\ref{equ2}) and then compared the continuum-corrected visibility with the continuum-corrected 
radiative transfer model of \citet{hillier01}. Figure~\ref{res_heicontsubline} (top) shows the measured 
flux and visibility for the MR-2004-12-26 measurement with the shortest projected baseline (43m) as well 
as the continuum-corrected visibility across the \ion{He}{I} emission component (solid red line). Because 
of the P Cygni-like absorption component, the continuum subtraction is highly uncertain in the blue 
region of the emission line (dashed red line in Fig.~\ref{res_heicontsubline}), as already discussed 
in the context of the Br$\gamma$ line in Sect.~\ref{sect:res_brg}. 

In Fig.~\ref{res_heicontsubline} (bottom), the continuum-corrected visibility of all AMBER data in the 
red region of the \ion{He}{I} emission line (averaged over the wavelength range 2.057--2.058$\,\mu$m) 
is shown as a function of spatial frequency. As the figure reveals, similar to the Br$\gamma$ emission, 
the visibilities inside the \ion{He}{I} emission line region reach rather low values. As the comparison 
shows, the line visibilities predicted by the model are much higher than the line visibilities measured 
with AMBER, indicating that the size of the line-emitting region in the model is too small. Rescaling 
of the model size by a factor of 2.4 results in a much better agreement between the model and observations 
(green curve in Fig.~\ref{res_heicontsubline}, bottom) and a FWHM diameter of $\gtrsim 8.2$~mas, which 
is 3.6 times larger than the FWHM diameter of 2.3~mas in the continuum. Due to the lack of interferometric 
data at small projected baselines, this value can only give a rough lower limit of the size.

The results for the visibility inside the \ion{He}{I} line can possibly be explained in a qualitative 
way in the framework of the binary model for the central object in $\eta$ Car 
\citep[e.g.][]{davidsonetal99,davidson01,pico02,hillier06,nielsen06}. In a model of this type, \ion{He}{I} 
emission should arise near the wind-wind interaction zone between the binary components. The hot secondary 
star is expected to ionize helium in a zone in the dense primary wind, adjoining the wind-wind interaction 
region. Such a region can produce \ion{He}{I} recombination emission.\footnote{For a qualitative sketch of 
the geometry, see ``zone 4'' in Figure~8 of \citet{martin06}, even though this figure was drawn to represent
He$^{++}$ in a different context.  For reasonable densities, the predicted He$^{+}$ zone has a 
quasi-paraboloidal morphology. In addition, some extremely dense cooled gas, labeled ``zone 6'' in the same 
figure, may also produce \ion{He}{I} emission.} The wind-wind shocked gas, by contrast, is too hot for
this purpose, while the density of the fast secondary wind is too low. Since the AMBER measurements (Dec.\ 
2004 and Feb.\ 2005, at orbital phases $\phi=0.268$ and $\phi=0.299$, see Table~\ref{tab:observations}) were 
obtained at an intermediate phase between periastron in July 2003 and apoastron in April 2006, the extension 
of the \ion{He}{I} emission zone is expected to be rather diffuse and larger than the continuum size. In other 
words, the \ion{He}{I} emission zone should be fairly extended and larger than the Hillier et al. model 
prediction, which is in agreement with the AMBER data.

\subsection{Differential Phases and Closure Phases}

The measurement of phase information is essential for the reconstruction of images from interferometric 
data, but such an image reconstruction is only possible with an appropriate coverage of the uv plane. 
Nevertheless, even single phase measurements, in particular of the closure phase and differential phase,
provide important information.

The closure phase (CP) is an excellent measure for asymmetries in the object brightness distribution. 
In our AMBER measurements, as illustrated in Figs.~\ref{fig:overview.BrG.ALL} and \ref{fig:overview.HeI.ALL}, 
we find that the CP in the \textit{continuum} is zero within the errors for all the various projected baselines 
of the UT2-UT3-UT4 baseline triplet, indicating a point-symmetric continuum object. However, in the line 
emission, we detect a non-zero CP signal in all data sets. In both MR measurements covering the Br$\gamma$ line, 
we find the strongest CP signal in the blue wing of the emission line at $\lambda=2.164\,\mu$m (-34\degr\ and 
-20\degr) and a slightly weaker CP signal in the red wing of the emission line at $\lambda=2.167\,\mu$m 
(+12\degr\ and +18\degr). We also detected non-zero CP signals in the HR measurement around Br$\gamma$ taken 
at a different epoch. In the case of the \ion{He}{I} line, a non-zero CP could only be detected at 
$\lambda=2.055\,\mu$m, just in the middle between the emission and absorption part of the P Cygni line profile. 

The differential phase (DP) at a certain wavelength bin is measured relative to the phase at all wavelength 
bins. Therefore, the DP measured within a wavelength bin containing line emission yields approximately the
Fourier phase of the combined object (continuum plus line emission) measured relative to the continuum. 
This Fourier phase might contain contributions from both the object phase of the combined object and a shift 
phase, which corresponds to the shift of the photocenter of the combined object relative to the photocenter
of the continuum object. Significant non-zero DPs were detected in the Doppler-broadened line wings of
the Br$\gamma$ line. Particularly within the blue-shifted wings, we found a strong signal (up to $\sim 
-60$\degr), whereas the signals are much weaker within the red-shifted line wings. These DPs might correspond 
to small photocenter shifts, possibly arising if the outer Br$\gamma$ wind region consists of many clumps which 
are distributed asymmetrically. The small differential phases of up to $\sim -15$\degr\ for the different
baselines of the blue-shifted light in the \ion{He}{I} line can perhaps also be explained by the above-mentioned 
asymmetries or within the framework of the binary model discussed in previous sections. In the binary model, 
a large fraction of the \ion{He}{I} is possibly emitted from the wind-wind collision zone, which is located 
between the primary and the secondary \citep{davidsonetal99,davidson01,pico02,hillier06,nielsen06}.

\subsection{Modeling with an inclined aspherical wind geometry}\label{brgmodel}

The goal of the modeling presented in this section is to find a model which is able to explain several 
remarkable features in our data; in particular, \textit{(a)}~the asymmetry in the Br$\gamma$ line profile 
(showing less emission in the blue-shifted wing than in the red-shifted wing) and the P Cygni-like 
absorption dip in the blue-shifted Br$\gamma$ wing, \textit{(b)}~the strong DP in the blue-shifted wing 
and a weaker DP signal in the red-shifted wing, and \textit{(c)}~the structure of the CP, showing a change
in the sign between the blue- and red-shifted line wing. We aimed for a geometrical but physically motivated 
model which would reproduce these features at all wavelength channels simultaneously. For this, we 
concentrate on the Br$\gamma$ line, as this line shows a stronger phase signal than the \ion{He}{I} line 
and was measured with a better uv coverage.

\begin{figure*}[t]
    \begin{minipage}[l]{42mm}
      \includegraphics[width=42mm,angle=0]{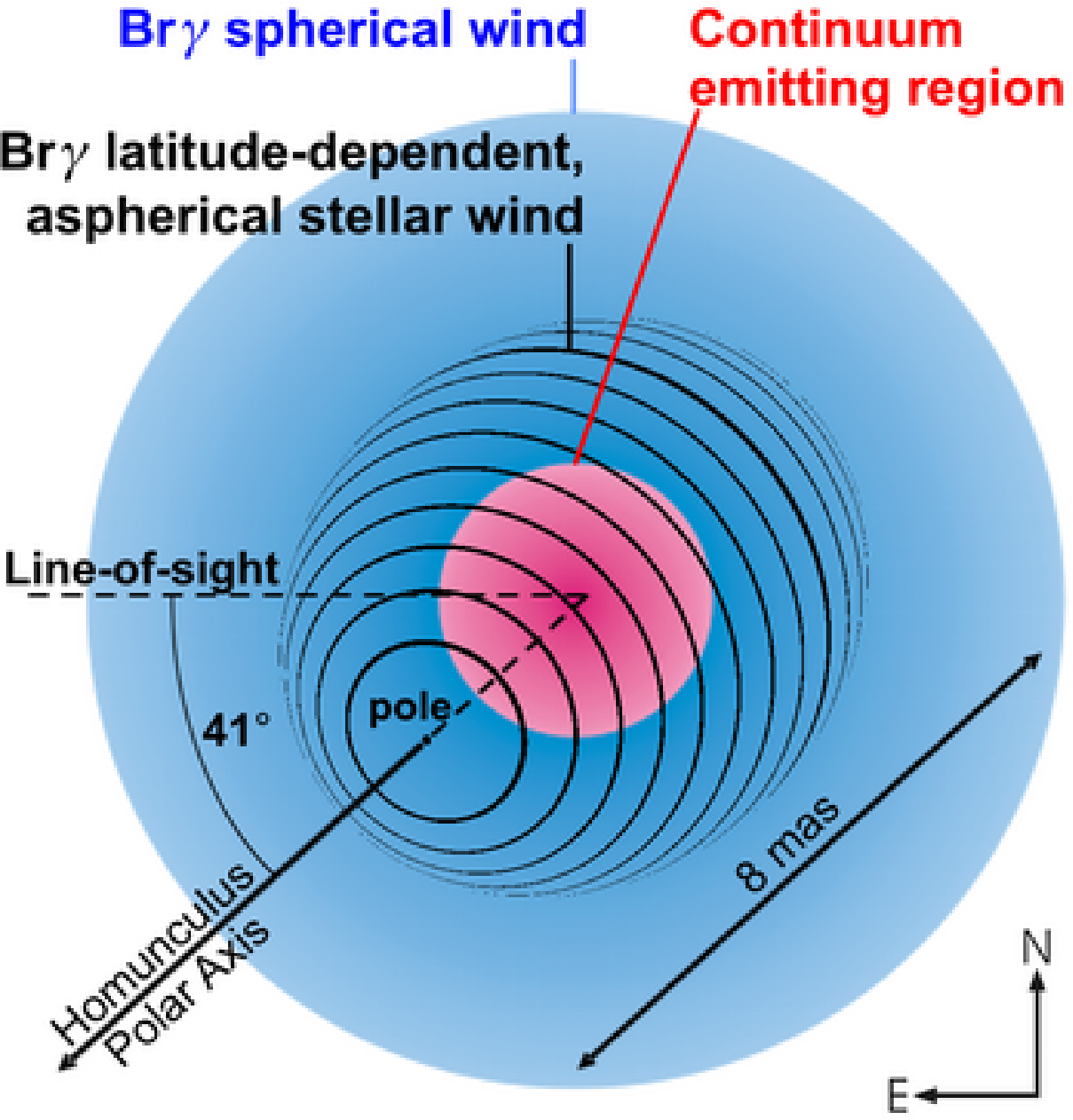}\\ 
      \caption{\label{fig:modelillu}
        \textbf{Left:} Illustration of the components of our geometric model for an optically thick, 
        latitude-dependent wind (see text for details). For the weak aspherical wind component, we 
        draw the lines of latitudes to illustrate the 3D-orientation of the ellipsoid.
        \textbf{Right (a, b):} The upper row shows the brightness distribution of the modeled 
        aspherical wind component (item (3) in the text) for two representative wavelengths. The 
        figures below show the total brightness distribution after adding the contributions from 
        the two spherical consituents of our model. 
      }
    \end{minipage}
    \begin{minipage}[l]{150mm}
      $\begin{array}{c @{\hspace{-16mm}} c}
        \textnormal{\textbf{a) Blue-shifted line wing}} &
        \textnormal{\textbf{b) Red-shifted line wing}}\\
        \lambda=2.1635\,\mu\textnormal{m} & \lambda=2.1669\,\mu\textnormal{m}\\[-12mm]
        \includegraphics[width=75mm,angle=0]{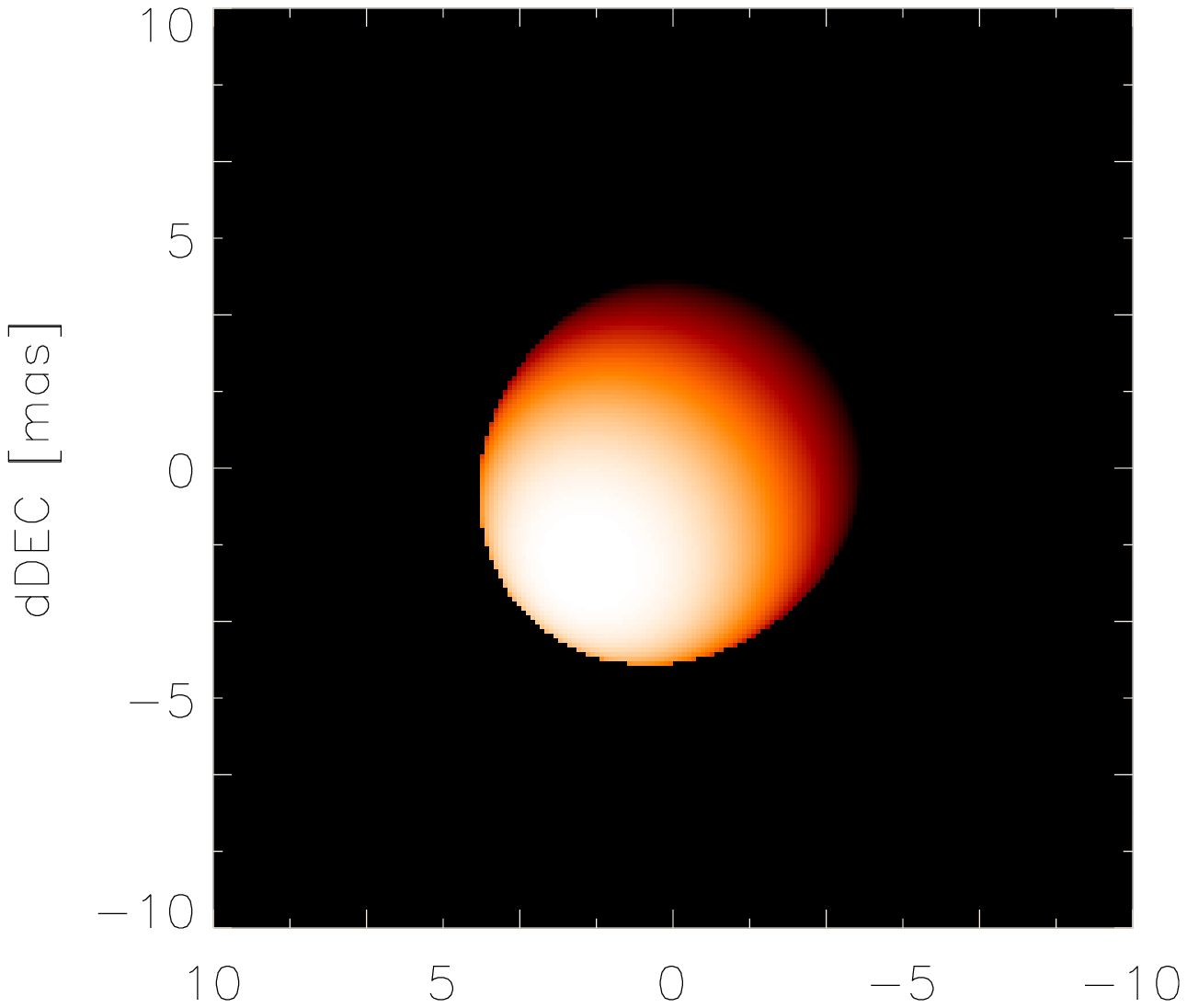} & 
        \includegraphics[width=75mm,angle=0]{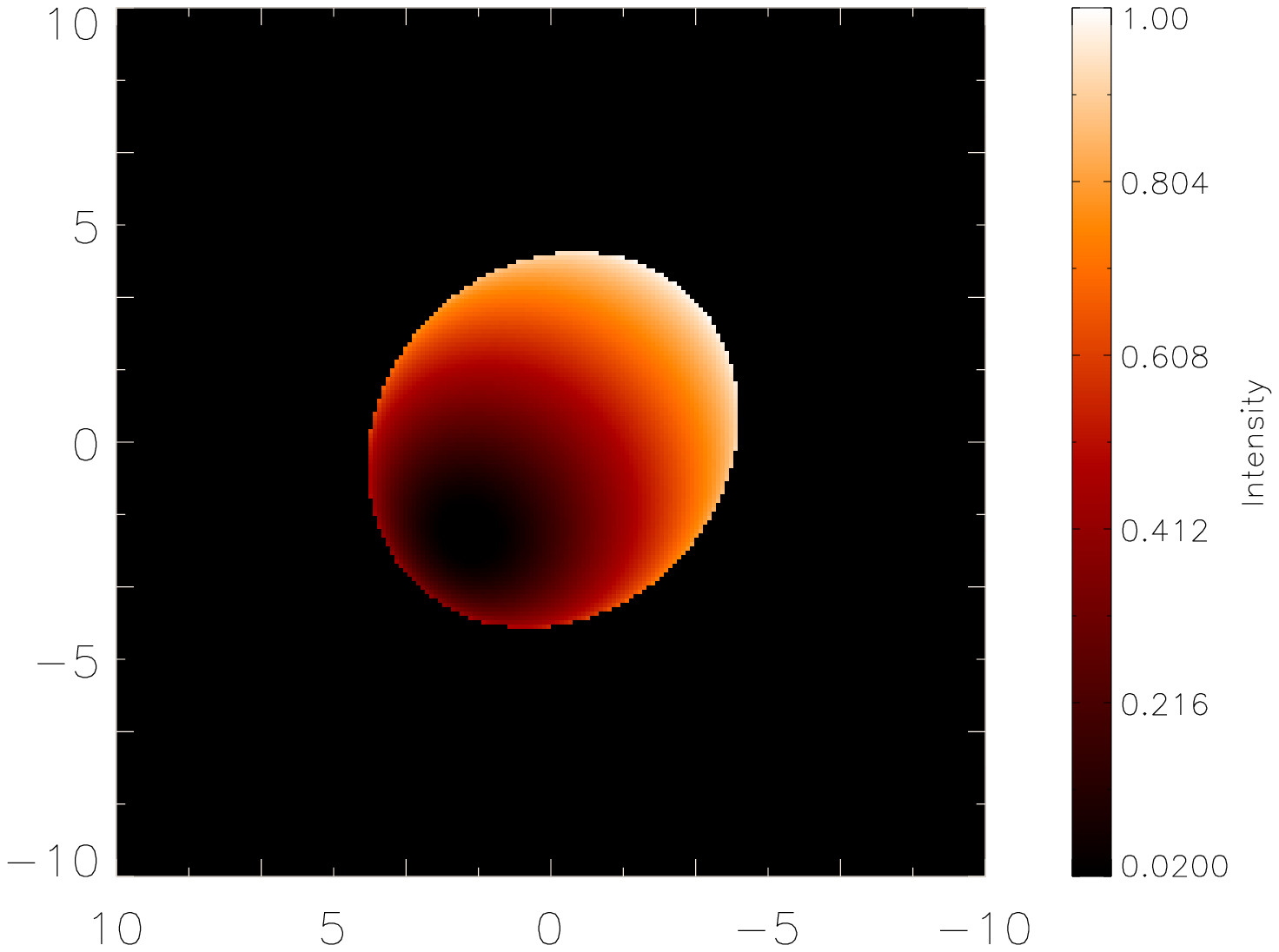} \\[-20mm]
        \includegraphics[width=75mm,angle=0]{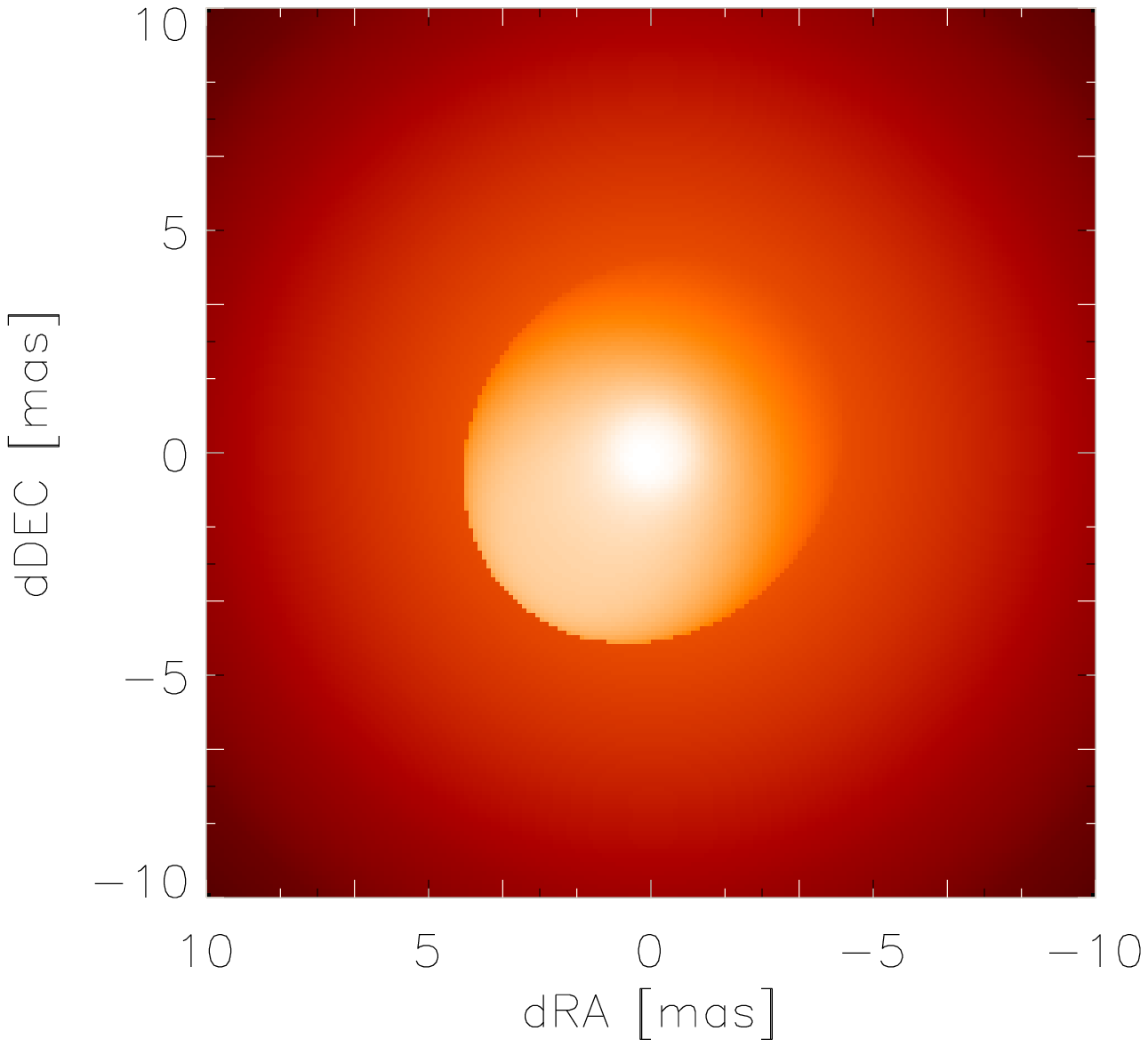} & 
        \includegraphics[width=75mm,angle=0]{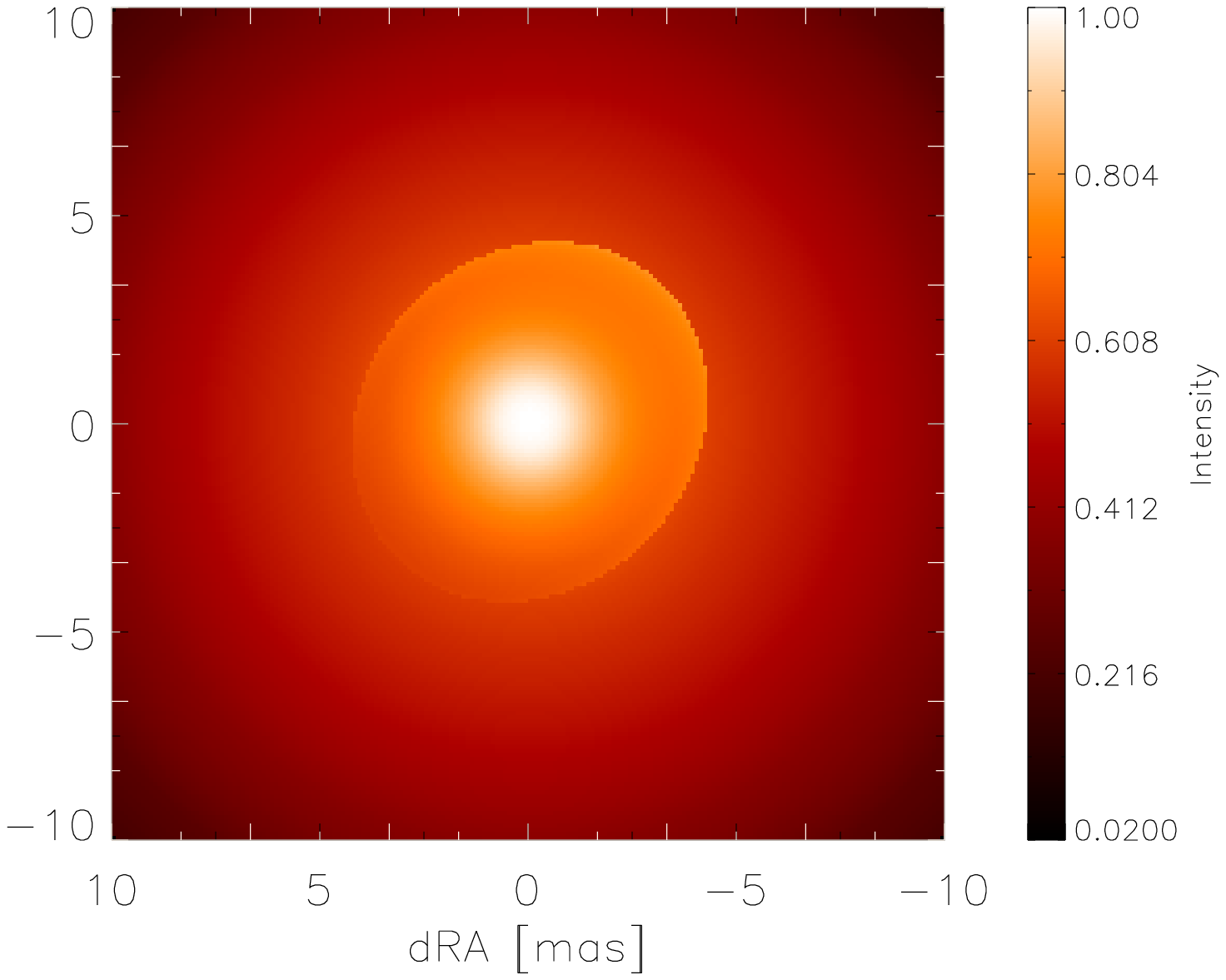} \\[-6mm]
      \end{array}$
      \vspace{15mm}
    \end{minipage}
    \vspace*{-7mm}
\end{figure*}

As \citet{smi03} convincingly showed, the stellar wind from $\eta$~Car seems to be strongly 
latitude-dependent, with the highest mass flux and velocities at the poles. This anisotropy 
can be understood in the context of theoretical models \citep[see, e.g.,][]{md01}, which take 
the higher temperatures at the poles ($g_{\textnormal{eff}}$-effect) and the equatorial gravity 
darkening on a rapidly rotating star into account \citep[\textit{von Zeipel} effect, ][]{zeipel25}.
As these models are quite successful in explaining the bipolar structure of the Homunculus nebula, 
we investigated whether such bipolar geometries with a latitude-dependent velocity distribution 
might also be suited to explain our interferometric data.

\begin{figure*}[p]
\centering
\includegraphics[width=160mm,angle=0]{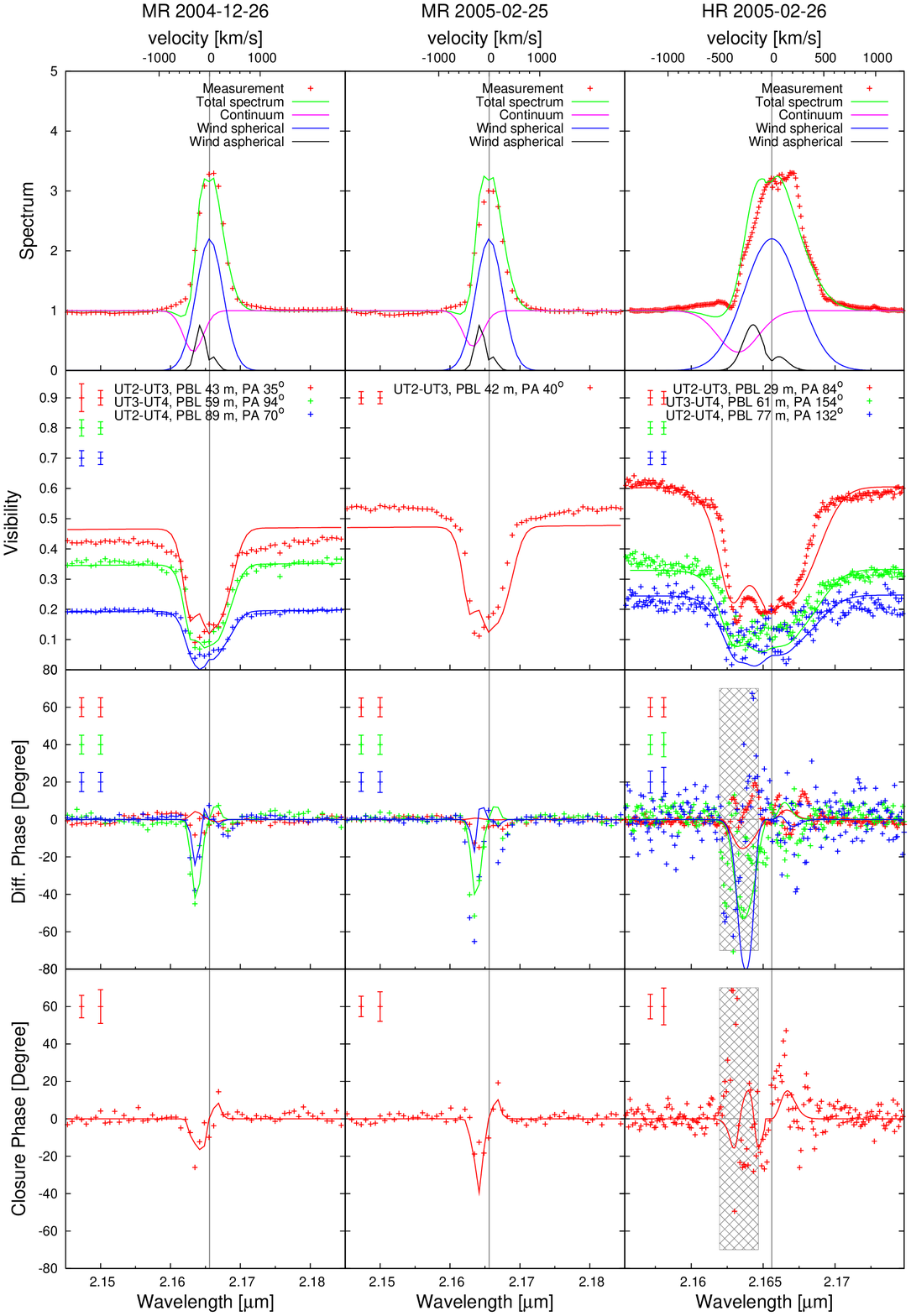}
\caption{\label{fig:model}
  Observables computed from our optically thick, latitude-dependent wind model (see 
  Fig.~\ref{fig:modelillu} for a model illustration). The points (crosses) represent the 
  measurements (as also shown in Fig.~\ref{fig:overview.BrG.ALL}), and the solid lines give 
  the observables computed from our model. The upper row shows the contributions from the 
  various model components to the total flux. Besides the continuum emission (purple line), 
  we introduced a spherical (blue line) and an aspherical (black line) wind component.
}
\end{figure*}

Due to its success in reproducing both the spectrum and the measured visibilities, we based our 
wind model on the spherical Hillier et al.\ (2001) model and superposed a weak aspherical stellar
wind geometry, which is inclined with respect to the line-of-sight. Our model includes three 
components (see Fig.~\ref{fig:modelillu}); namely, 
\begin{description}
\item[(1)] a continuum component (using the Hillier et al.\ continuum CLV, see
           Fig.~\ref{res_nlte_visclv} top) with a blue-shifted absorption component,
\item[(2)] a spherical stellar wind (using the Hillier et al.\
           continuum-subtracted Br$\gamma$ CLV), and
\item[(3)] an aspherical wind geometry, represented by a 41\degr\ inclined ellipsoid.
\end{description}
The relative contribution of these different constituents to the total flux is given by the input 
spectra shown in the upper row of Fig.~\ref{fig:model}. For the spherical and aspherical wind 
component, we assume Gaussian-shaped spectra. The original Hillier et al.\ CLVs slightly underestimate 
the size of the observed structures (see Fig.~\ref{res_nlte_vislam}). Therefore, we rescaled them by 
10\% to obtain a better agreement for the visibilities at continuum wavelengths.

The aspherical wind of $\eta$~Car is simulated as an ellipsoid with an inclination similar to the 
inclination angle of the Homunculus \citep[41\degr,][]{smi06}. While the south-eastern pole (which 
is inclined towards the observer) is in sight, the north-western pole is obscured. The 
latitude-dependent velocity distribution expected for the $\eta$~Car wind was included in our model 
by coupling the latitude-dependent brightness distribution of the ellipsoid to the wind velocity. 
At the highest \textit{blue-shifted} velocities, mainly the \textit{south-eastern polar region} 
contributes to the emission (see Fig.~\ref{fig:modelillu}a). In the \textit{red-shifted} line
wing, mainly the (obscured) \textit{north-western pole} radiates (see Fig.~\ref{fig:modelillu}b). 
The axis of the ellipsoid was assumed to be oriented along the Homunculus polar axis (PA~132\degr, 
\citealt{smi06}) and its axis ratio was fixed to 1.5.

As our simulations show, such an asymmetric geometry can already explain the measured DPs and CPs 
with a rather small contribution of the asymmetric structure to the total flux (see black line in 
Fig.~\ref{fig:model}, upper row). Although the large number of free parameters prevented us from 
scanning the whole parameter space, we found reasonable agreement with a size of the ellipsoid 
major axis of 8~mas. Fig.~\ref{fig:model} shows the spectrum, visibilities, DPs, and CPs computed 
from the model.

As our model was inspired by physical models, but does not take the
complicated radiation transport and hydrodynamics involved in reality into
account, we would like to note that our model allows us to check for consistency
between the considered geometry and the AMBER spectro-interferometric data,
but can neither constrain the precise parameters of a possible aspherical
latitude-dependent stellar wind around $\eta$~Car, nor can it rule out other
geometries.  We summarize some qualitative properties of our wind model as
follows:
\begin{itemize}
\item The strong underlying spherical component mainly accounts for the very
  low visibilities measured within the line.
\item The aspherical wind component introduces the asymmetry required to
  roughly explain the measured phase signals.  In particular, it
  reproduces the larger DPs and CPs within the blue-shifted line
  wing compared to the red-shifted wing, as the red-shifted emission region
  is considerably obscured. It also accounts for the flip in the CP sign,
  as the photocenter of the line emission shifts its location between the
  blue- and red-shifted wing relative to the continuum photocenter.
\item The absorption component which we introduced in the blue-shifted wing of
  the Br$\gamma$ line allows us to reproduce the asymmetry measured in the
  shape of the Br$\gamma$ emission line profile (showing an increase of
  flux towards red-shifted wavelengths) and the weak dip observed at
  far-blue-shifted wavelengths.  Furthermore, with the decrease of the
  continuum contribution, the absorption component helps to lower the
  visibilities in the blue-shifted line wing, simultaneously increasing
  the asymmetry in the brightness distribution (increasing the phase
  signals).  Finally, with the interplay between the absorption and
  emission component, our simulation reproduced a "bump" in the visibility
  similar to the one observed on the shortest baseline of our HR
  measurement ($\lambda \approx 2.163 \mu$m).
\end{itemize}

\subsection{Feasibility of the detection of the hypothetical hot companion and the
  wind-wind interaction zone}

\begin{figure*}[t]
  $\begin{array}{ccc}
    \begin{minipage}[l]{65mm}
      $\begin{array}{c}
        \textnormal{\textbf{a) Continuum emission with binary}}\\[-8mm]
        \includegraphics[width=55mm,angle=-0]{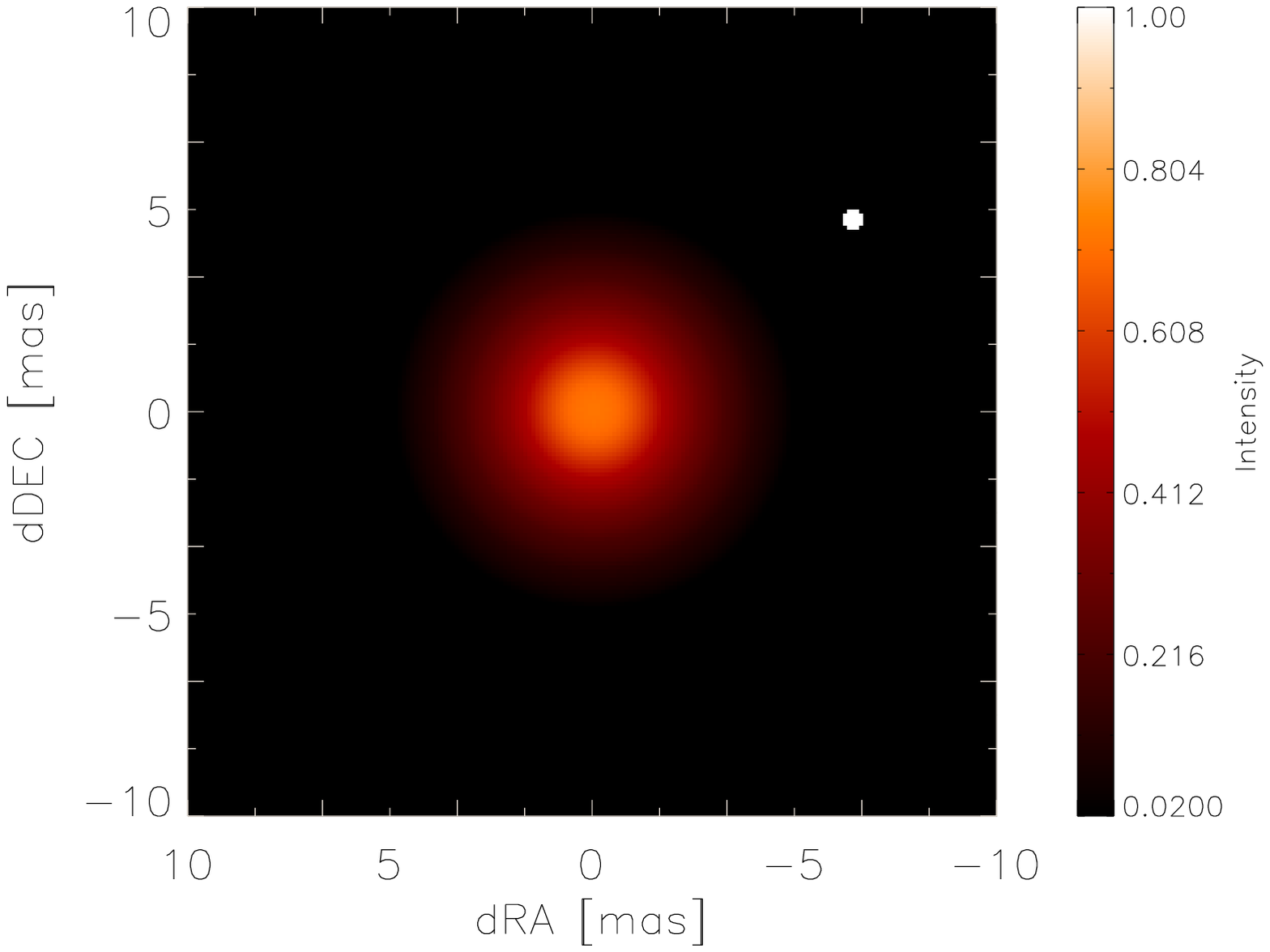}\\[-2mm]
        \textnormal{\textbf{b) 2D-Visibility without binary}}\\[-8mm]
        \includegraphics[width=55mm,angle=0]{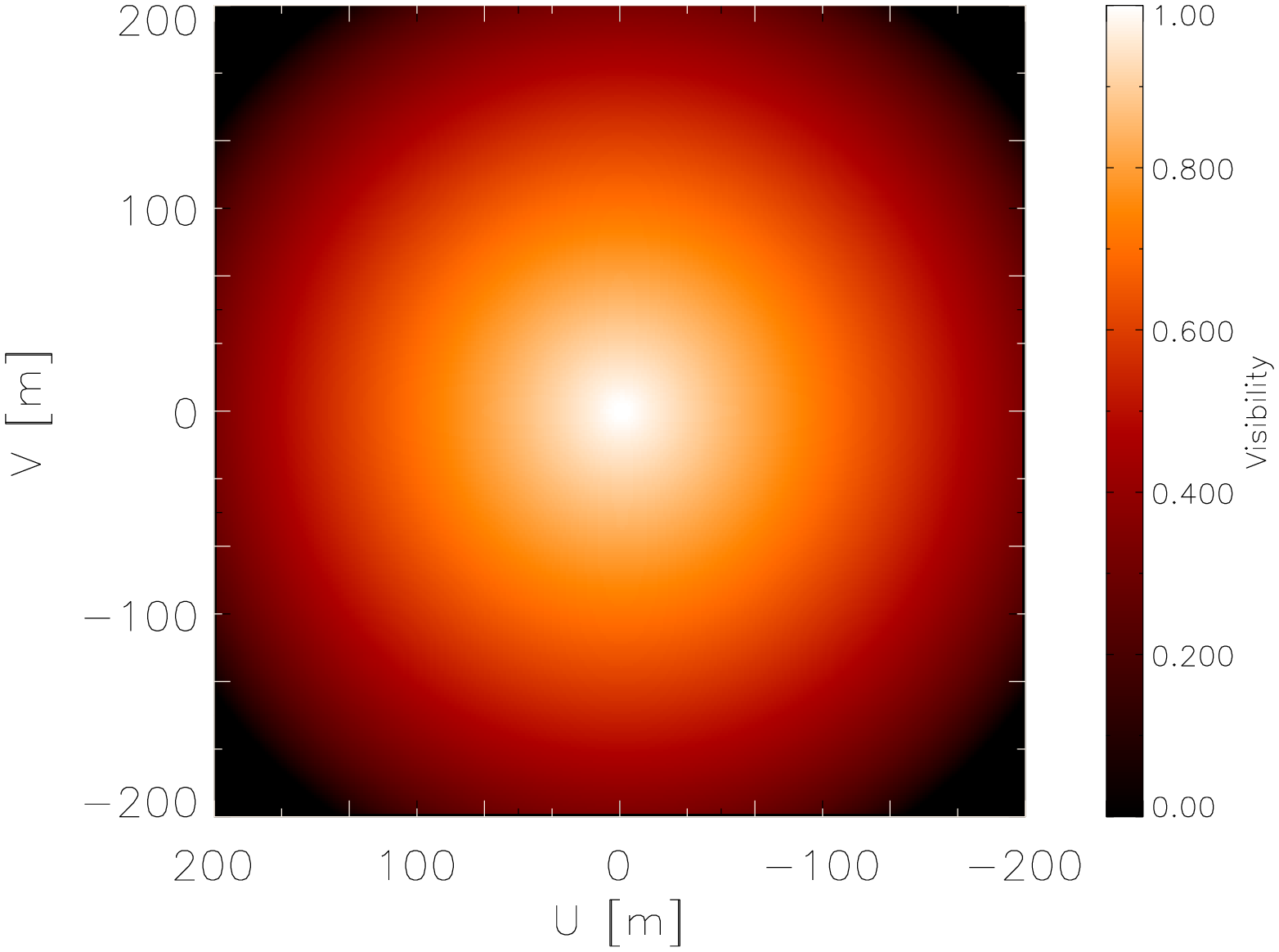}\\[-2mm]
        \textnormal{\textbf{c) 2D-Visibility with binary}}\\[-8mm]
        \includegraphics[width=55mm,angle=0]{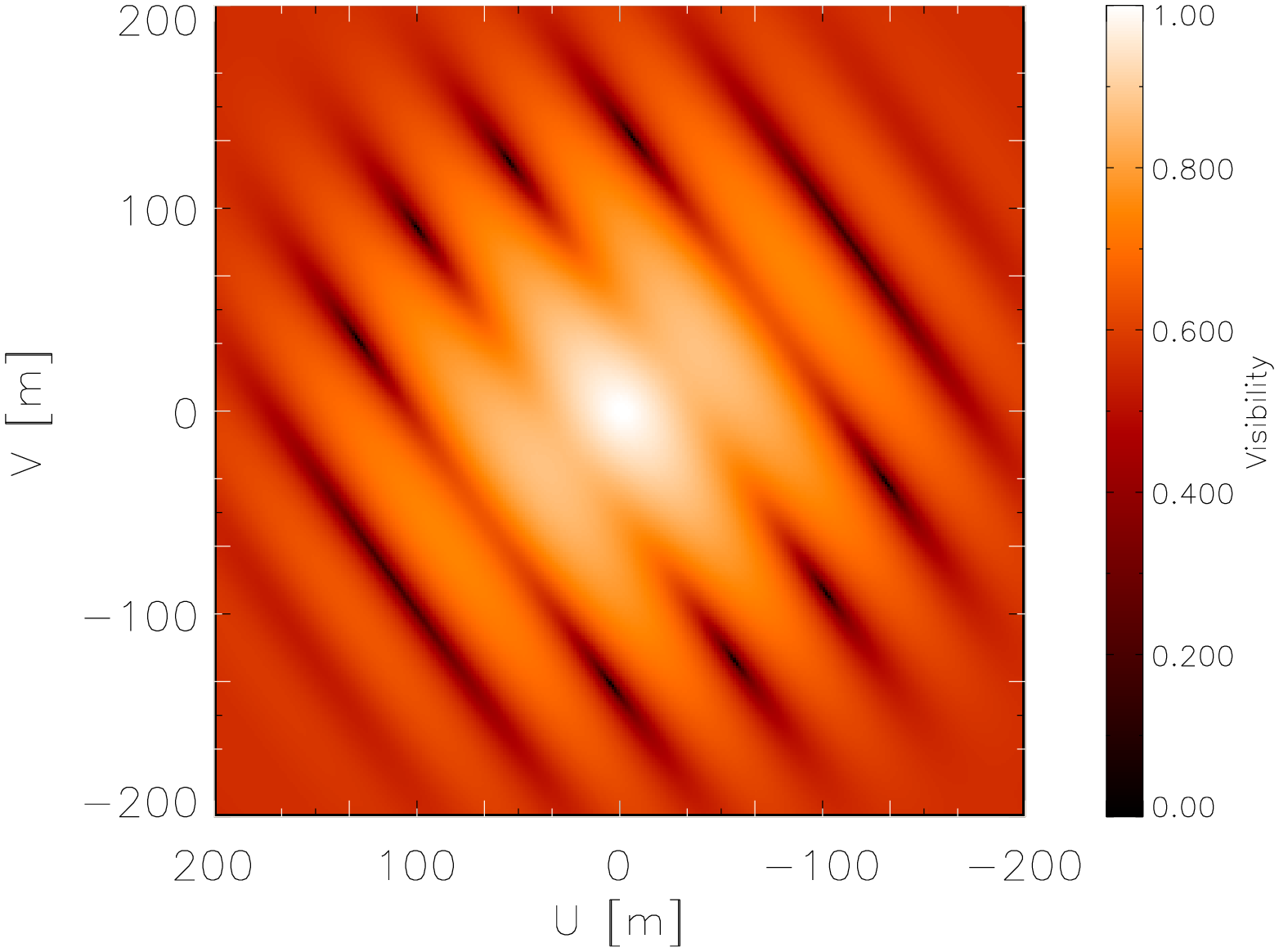}
      \end{array}$
    \end{minipage} &

    \begin{minipage}[l]{110mm}
      $\begin{array}{c}
        \textnormal{\textbf{d) Residuals between models with/without a binary}}\\[5mm]
        \includegraphics[width=110mm,angle=0]{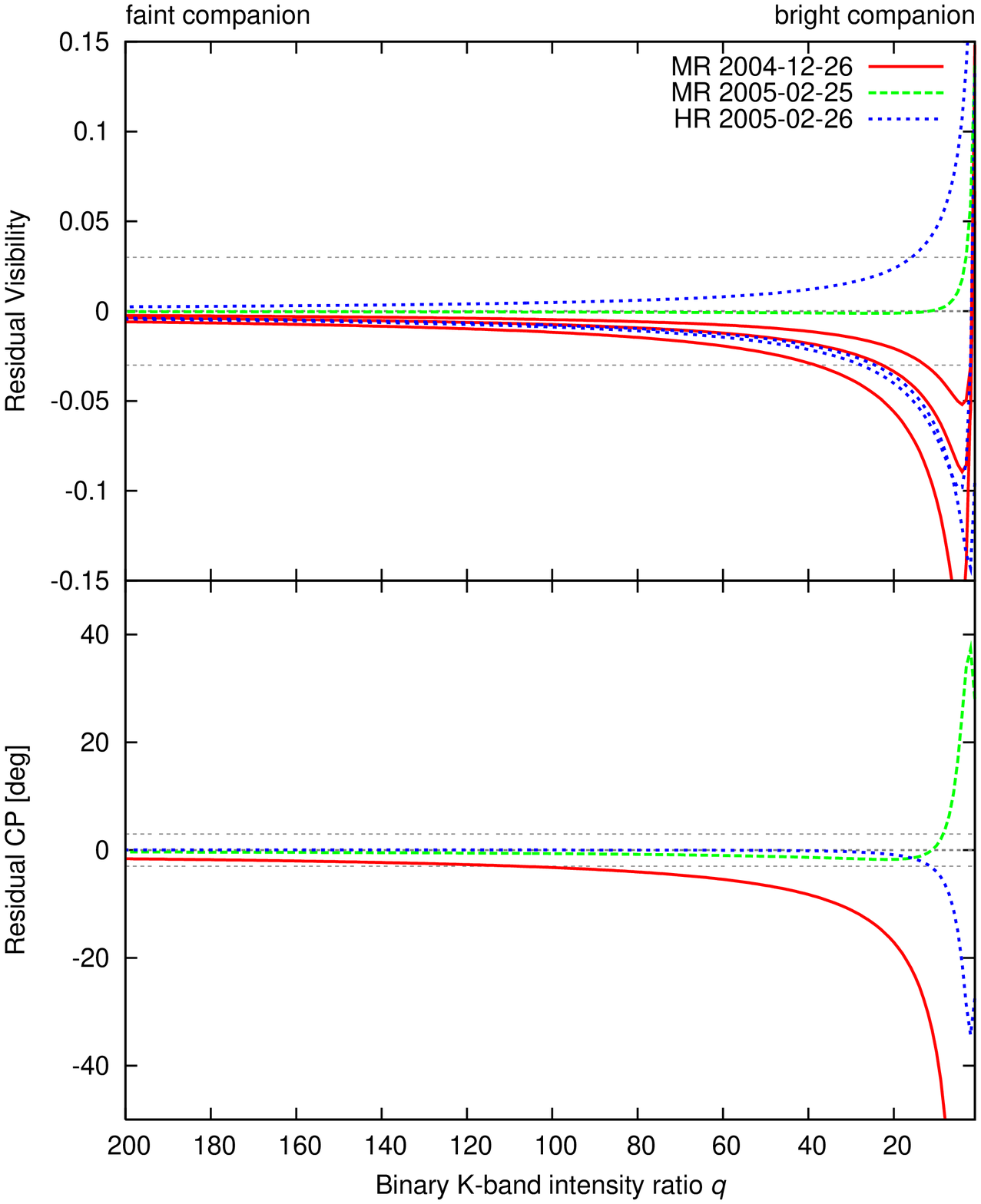}
      \end{array}$
    \end{minipage} &
  \end{array}$

      \caption{\label{fig:binaryresiduals}
        Simulation illustrating the signatures of a binary companion at the predicted
        position \citep[for the orbital phase at the time of our continuum observations around the 
        Br$\gamma$ line; separation $\sim8$~mas, $PA \sim-36\degr$; see][]{nielsen06}.
        For the primary component, we assume a continuum emission CLV from \citet{hillier06}, 
        to which we add a uniform disk of 0.1~mas diameter to account for the binary companion 
        (see \textbf{a}). Figure \textbf{b} shows the 2D-visibility function of a single component 
        model containing only the continuum-emitting object with Hillier-type CLV. Figure \textbf{c} 
        shows the 2D-visibility corresponding to the binary model shown in panel \textbf{a} for a 
        small $K$-band flux ratio $q=10$ between the continuum emitting region (Hillier-type CLV) 
        and the companion. In \textbf{d} the residuals of the visibilities (upper panel) and closure 
        phases (lower panel) between the models with and without a companion are shown as a function 
        of flux ratio $q$ for baselines and PAs corresponding to the seven AMBER measurements. The 
        dashed horizontal lines indicate the uncertainty of the measurements. Given these uncertainties, 
        the figure illustrates that a companion signature should be visibile from the AMBER closure 
        phase measurements, if the binary is less than $\sim110$ times fainter compared to the primary. 
        See text for details.
      }
\label{phases:binary}
\end{figure*}

One of the most intriguing questions regarding $\eta$~Car is whether or not its central object is a 
binary, as suggested to explain 
cycle \citep[e.g., ][]{dam96}.

\subsubsection{A simple binary continuum model}

To investigate whether the AMBER measurements presented here can shed more light on the binarity hypothesis, 
we used the following approach: We constructed a {\it simple binary model} consisting of a primary wind 
component with a CLV according to the continuum model of Hillier et al.\ (2006; FWHM diameter $\sim2.3\,$mas; 
see upper panels in Fig.~\ref{res_nlte_visclv}) and an unresolved binary companion represented by a point-like 
source (uniform disk with $\sim0.1$~mas FWHM diameter). The secondary component is predicted to be approximately 
located at PA --36\degr\ with a separation of 8~mas from the primary for the time of the AMBER observations 
\citep{nielsen06}. The continuum flux ratio $q$ was treated as a free parameter. We would like to note that
in our model, we assumed that all $K$-band light from the secondary is reaching us unprocessed; i.e.\ we 
ignored a possible dilution or re-distribution of the secondary's radiation.

We calculated the 2D visibility function (see Fig.~\ref{phases:binary}c) of this model intensity distribution
for different values of $q$, as well as the closure phases for the baselines and PAs corresponding to our 
AMBER measurements (Fig.~\ref{phases:binary}d). Finally, we compared the results with those obtained from a 
single component model where only the primary wind is present (Fig.~\ref{phases:binary}b). The differences of 
the visibilities and closure phases between the single star and the binary model (at the baselines and PAs 
corresponding to our AMBER measurements) are displayed in Fig.~\ref{phases:binary}d as a function of the 
$K$-band flux ratio of the binary components.

Fig.~\ref{phases:binary}d shows two interesting results: First, the closure phase is more sensitive 
to the binary signature than the visibilities and, thus, a more suitable observable to constrain the 
binary hypothesis. And second, given the accuracies of our first AMBER visibility and closure phase 
measurements (indicated by the horizontal dashed-dotted lines), we can conclude, for the particular 
model shown in Fig.~\ref{phases:binary}, that the AMBER closure phases put an intensity ratio limit 
$q_{\rm min} \approx 110$ on the binary $K$-band flux ratio. This limit is in line with the estimate 
$q\approx 200$ given by \citet{hillier06}. Thus, based on the model shown in Fig.~\ref{phases:binary}, 
the AMBER measurements are not in conflict with recent model predictions for the binary. 

To investigate whether we can put similar constraints on the minimum $K$-band flux ratio $q_{\rm min}$ 
for arbitrary separations and PAs, we calculated a larger grid of binary models and compared the residuals
of visibilities and closure phases analogue to the example shown in Fig.~\ref{phases:binary}.

For these grid calculations, we used values in the range from 4 to 14~mas for the binary separation with 
increments of 1~mas, and PAs of the secondary in the whole range from 0\degr\, to 360\degr\, in steps of 
10\degr. The $K$-band flux ratio $q$ of the binary components was varied in the range from 1 to 250 with 
$\Delta q=4$. As a result of the grid calculation, we obtained the minimum $K$-band flux ratio $q_{\rm min}$ 
as a function of binary separation and orientation.

Whereas the study with a fixed companion position presented above allowed us to put rather stringent 
constraints on $q_{\rm min}$ (see Fig.~\ref{phases:binary}), this systematic study revealed that due 
to the rather poor uv coverage, a few very specific binary parameter sets exist where we are only 
sensitive to $q_{\rm min} \approx 10$. Nevertheless, for the above-mentioned separation interval 
(4 to 14~mas), we found that we are able to detect companions up to $q_{\rm min}=50$ at more than 90\%
of all PAs. In order to push this sensitivity limit in future observations, a better uv coverage will 
be required. Together with the expected higher closure phase accuracy, AMBER will be sensitive up to 
$q> 200$ and, therefore, have the potential to probe the currently favored binary models.

\subsubsection{Can AMBER detect a \ion{He}{I} wind-wind interaction zone shifted a few mas from the 
primary wind?}\label{amber_performance}

In the context of the binary hypothesis, it is also important to discuss the implications for the interpretation 
of the AMBER \ion{He}{I} measurements. According to the binary model, a large fraction of the \ion{He}{I} line 
emission should arise from the wind-wind collision zone expected between the primary and the secondary 
\citep{davidsonetal99,davidson01,hillier06,nielsen06}. The exact intensity ratio of primary \ion{He}{I} wind 
and \ion{He}{I} emission from the wind-wind interaction zone is not known. Figure ~\ref{res_nlte_vislam} 
suggests that during the AMBER observations, the total \ion{He}{I} flux was roughly two times larger than the 
model prediction of \citet{hillier01} for the primary \ion{He}{I} wind.

At the orbital phases of the AMBER measurements, the wind-wind collision zone should be at resolvable distances 
from $\eta$~Car's primary \citep[resolution $\sim5\,$ mas; companion separation $\sim8\,$mas, PA 
$\sim$\,--36\degr, ][]{nielsen06}. Looking at the AMBER \ion{He}{I} data, we see that the differential as well 
as the closure phases are zero everywhere except for the transition region between the absorption and emission 
part of the \ion{He}{I} line, where we find differential phases of $\sim10$--20\degr\, and a closure phase of
$\sim$\,--30\degr; i.e., the phases measured across the \ion{He}{I} line are significantly weaker compared
to the Br$\gamma$ line. The question is now, why AMBER measured weaker phase signals within the \ion{He}{I} 
line and if this result is in line with the predictions of the wind-wind collision model.

One possible explanation for the small measured phases could be the orientation of the binary orbit.
If the orbit's major axis is nearly aligned with the line-of-sight, the photocenter shift inside 
the \ion{He}{I} line will be very small. In addition, the deviations from point symmetry would be rather
small. Therefore, in the case of this special geometry, both differential phases and the closure phase
would be small, in qualitative agreement with the AMBER data. Another explanation could be that the 
contribution of the wind-wind collision zone to the \ion{He}{I} line emission is much weaker than that 
of the primary wind. However, this is not very likely \citep[see][]{hillier06}.

A different explanation for the weak phases can be found from a modeling approach similar to the one for 
the Br$\gamma$ line region outlined in Sect.~\ref{brgmodel}.  Based on the results presented in 
Sects.~\ref{res:overview} and \ref{sect:res_hei}, we constructed a simple \ion{He}{I} model consisting of 
a spherical primary wind component with a Hillier-type CLV (2.5~mas FWHM diameter) and an extended spherical 
\ion{He}{I} line-emitting region with Gaussian CLV and a 7~mas FWHM diameter (i.e., for simplicity, we assumed 
that all \ion{He}{I} flux is emitted from the wind-wind interaction region; however, some fraction of 
\ion{He}{I} is also emitted from the primary wind; see Hillier et al.\ 2006 and Fig.~\ref{res_nlte_vislam} 
of the present paper). The center of the line-emitting component of this model is located 3~mas away from the 
primary wind component towards PA 132\degr; i.e., in the direction of the Homunculus axis. The spectra of the 
continuum and line-emitting components were chosen in such a way that the combined spectrum resembles the 
observed \ion{He}{I} line spectrum.

The modeling results show that this simple model is approximately able to simultaneously reproduce the observed
spectrum and the wavelength dependence of visibilities, differential phases (10--20\degr), and closure phases 
($\sim$\,--30\degr). Thus, our simple model example illustrates that the AMBER measurements can be understood 
in the context of a binary model for $\eta$~Car and the predicted \ion{He}{I} wind-wind collision scenario 
\citep[e.g.][]{davidsonetal99,davidson01,hillier06,nielsen06}. We note that the model parameter values given 
above are of preliminary nature. A more detailed, quantitative modeling is in preparation and will be subject 
of a forthcoming paper. Furthermore, we would like to emphasize, as already discussed in previous sections, 
that there are likely to be {\it three} sources of \ion{He}{I} emission - the primary wind, a wind-wind 
interaction zone (bow shock), and the ionized wind zone caused by the ionization of the secondary. For both 
the bow shock and the ionized wind zone, the ionizing UV radiation field of the secondary is of crucial 
importance. On the basis of the observed blue-shift and the weakness of the \ion{He}{I} during the event, 
we believe that the primary wind contribution is small. It is not yet possible to decide on the relative 
contributions of the bow shock and the ionized wind region.

\section{Conclusions}
In this paper we present the first near-infrared spectro-interferometry of the enigmatic Luminous Blue Variable
$\eta$ Car obtained with AMBER, the 3-telescope beam combiner of ESO's VLTI. In total, three measurements with 
spectral resolutions of $R=1,500$ and $R=12,000$ were carried out in Dec.\ 2004 ($\phi=0.268$) and Feb.\ 2005 
($\phi=0.299$), covering two spectral windows around the \ion{He}{I} and Br$\gamma$ emission lines at $\lambda=2.059$ 
and $2.166\,\mu$m, respectively. From the measurements, we obtained spectra, visibilities, differential visibilities, 
differential phases, and closure phases. From the analysis of the data, we derived the following conclusions:
\begin{itemize}
\item In the $K$-band continuum, we resolved $\eta$ Car's optically thick wind. From a Gaussian fit of the $K$-band 
      continuum visibilities in the projected baseline range from 28--89~m, we obtained a FWHM diameter of $4.0\pm0.2$ 
      mas. Taking the different fields-of-view into account, we found good agreement between the AMBER measurements 
      and previous VLTI/VINCI observations of $\eta$ Car presented by \citet{van03}.
\item When comparing the AMBER {\it continuum} visibilities with the NLTE radiative transfer model from \citet{hillier01}, 
      we find very good agreement between the model and observations. The best fit was obtained with a slightly rescaled
      version of the original Hillier et al.\ model (rescaling by 1--2\%), corresponding to FWHM diameters of 2.27 mas 
      at $\lambda=2.040\,\mu$m and 2.33~mas at $\lambda=2.174\,\mu$m.
\item If we fit \citet{hillier01} model visibilities to the observed AMBER visibilities, we obtain, for example, 50\% 
      encircled-energy diameters of 4.2, 6.5, and 9.6~mas in the 2.17$\,\mu$m continuum, the \ion{He}{I}, and the 
      Br$\gamma$ emission lines, respectively.
\item In the continuum around the Br$\gamma$ line, we found an asymmetry towards position angle PA=$120\degr\pm15\degr$ 
      with a projected axis ratio of $\xi=1.18\pm0.10$. This result confirms the earlier finding of \citet{van03} using 
      VLTI/VINCI and supports theoretical studies which predict an enhanced mass loss in polar direction for massive 
      stars rotating close to their critical rotation rate \citep[e.g.\ ][]{owo96,owo98}.
\item For both the Br$\gamma$ and the \ion{He}{I} emission lines, we measured non-zero differential phases and non-zero 
      closure phases within the emission lines, indicating a complex, asymmetric object structure.
\item We presented a physically motivated model which shows that the asymmetries (DPs and CPs) measured within the wings 
      of the {\it Br$\gamma$ line} are consistent with the geometry expected for an aspherical, latitude-dependent 
      stellar wind.  Additional VLTI/AMBER measurements and radiative transfer modeling will be required to determine
      the precise parameters of such an inclined aspherical wind.
\item Using a simple binary model, we finally looked for a possible binary signature in the AMBER closures phases. 
      For separations in the range from 4 to 14~mas and arbitrary PAs, our simple model reveals a minimum $K$-band 
      flux ratio of $\sim$50 with a 90\% likelihood.
\end{itemize}

Our observations demonstrate the potential of VLTI/AMBER observations to unveil new structures of $\eta$~Car on the 
scales of milliarcseconds. Repeated observations will allow us to trace changes in observed morphology over $\eta$~Car's 
spectroscopic 5.5~yr period, possibly revealing the motion of the wind-wind collision zone as predicted by the 
$\eta$~Car binary model. Furthermore, future AMBER observations with higher accuracy might be sensitive enough to 
directly detect the hypothetical hot companion.

\begin{acknowledgements}
   This work is based on observations made with the European Southern Observatory telescopes. This research has 
   also made use of the ASPRO observation preparation tool from the \emph{Jean-Marie Mariotti Center} in France, 
   the SIMBAD database at CDS, Strasbourg (France) and the Smithsonian/NASA Astrophysics Data System (ADS).

   We thank the referee Dr.~N.~Smith and Dr.~A.~Damineli for very valuable comments and suggestions
   which helped to considerably improve the manuscript.

   The data reduction software \texttt{amdlib} is freely available on the AMBER site 
   \texttt{http://amber.obs.ujf-grenoble.fr}. It has been linked with the free software Yorick
   \footnote{\texttt{ftp://ftp-icf.llnl.gov/pub/Yorick}} to provide the user-friendly interface 
   \texttt{ammyorick}.

   The NSO/Kitt Peak FTS data used here to identify the telluric lines in the AMBER data were produced 
   by NSF/NOAO. 

   This project has benefitted from funding from the French Centre National de la Recherche Scientifique (CNRS) 
   through the Institut National des Sciences de l'Univers (INSU) and its Programmes Nationaux (ASHRA, PNPS). 
   The authors from the French laboratories would like to thank the successive directors of the INSU/CNRS
   directors. S.~Kraus was supported for this research through a fellowship from the International Max Planck 
   Research School (IMPRS) for Radio and Infrared Astronomy at the University of Bonn. C.~Gil's work was partially 
   supported by the Funda\c{c}\~ao para a Ci\^encia e a Tecnologia through project POCTI/CTE-AST/55691/2004 from 
   POCTI, with funds from the European program FEDER. K.~Weis is supported by the state of North-Rhine-Westphalia 
   (Lise-Meitner fellowship).
\end{acknowledgements}

\bibliographystyle{aa}
\bibliography{5577}

\appendix
\section{Wavelength calibration}\label{app:wavecal}

\begin{figure}[t]
  \centering
  \includegraphics[width=85mm,angle=0]{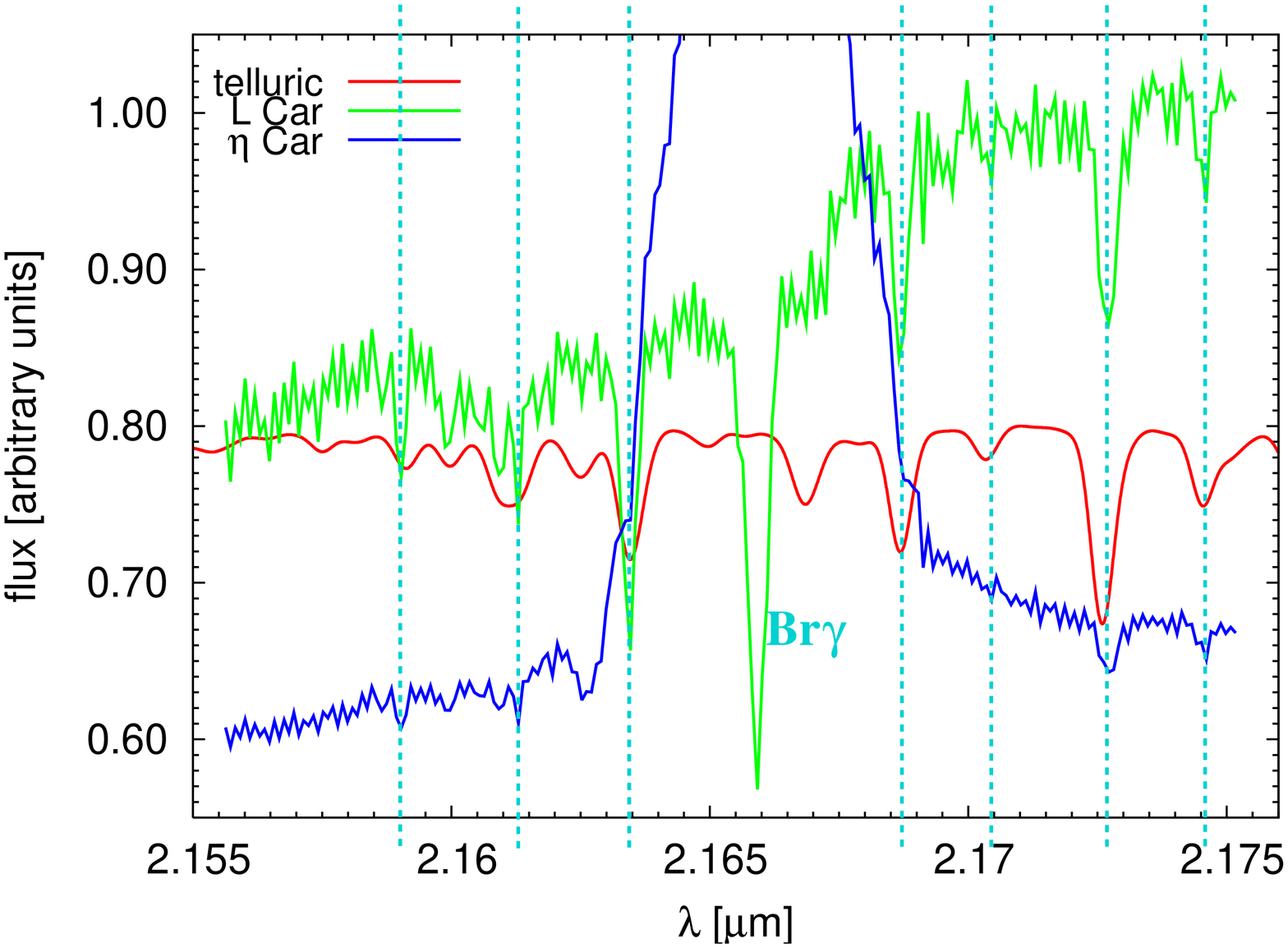}
\vspace*{-3mm}

  \includegraphics[width=58mm,angle=-90]{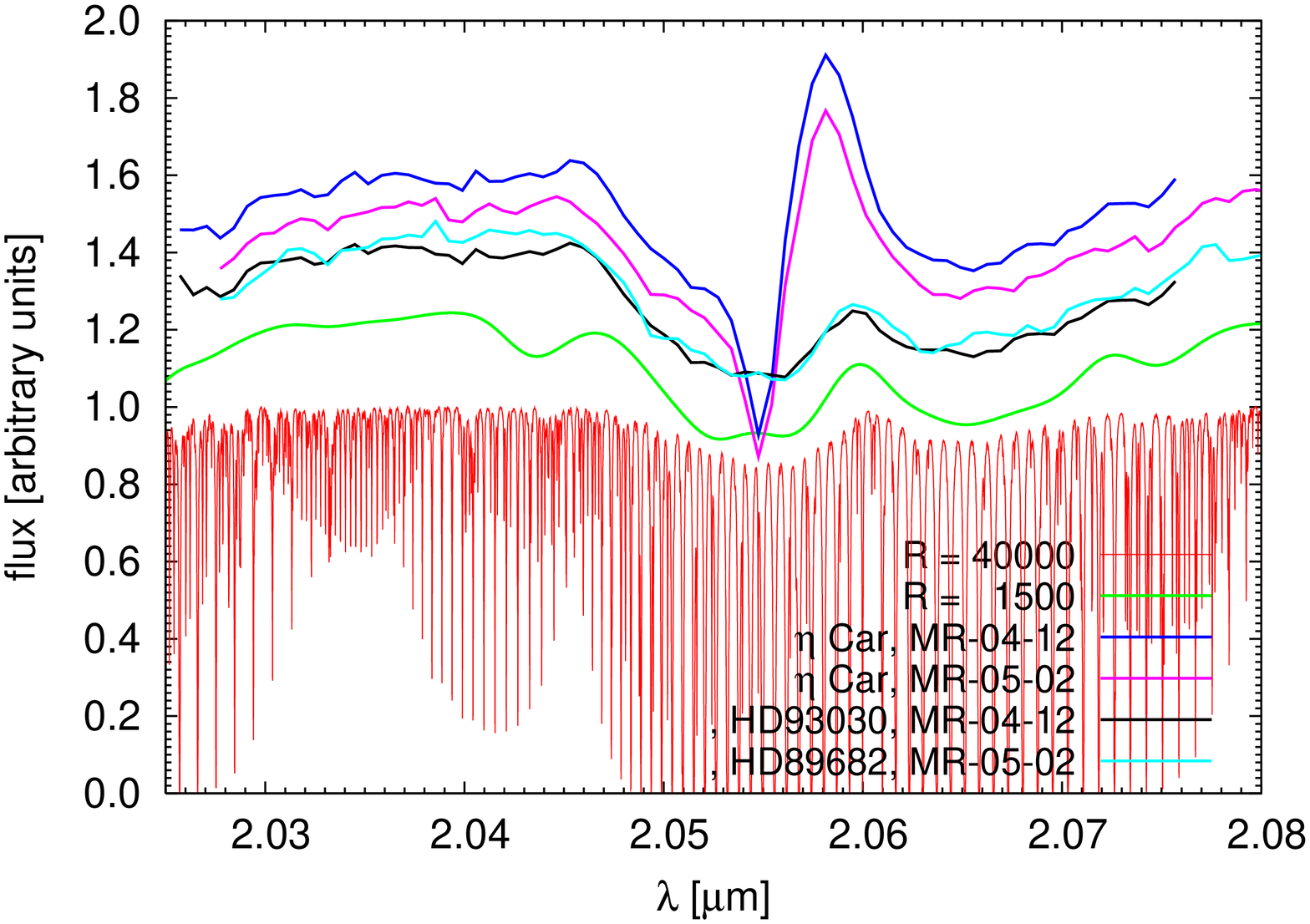}
  \caption{
{\bf Top}:
Spectral calibration of the AMBER observations. The figure shows the spectrum of $\eta$ Car (blue) and 
the calibrator star L~Car (green) around the Br$\gamma$ line obtained from the HR measurements in Feb.\ 2005. 
In addition, a telluric spectrum from the Kitt Peak Observatory is shown in red. The original telluric spectrum
with a spectral resolution of $R=40,000$ was spectrally convolved to match the resolution of the AMBER measurements. 
For the flux calibration of the $\eta$ Car spectrum, the Br$\gamma$ line in the L Car spectrum was interpolated 
before the division. See text for further details.
{\bf Bottom}:
Same as top panel, but for both MR measurements around the \ion{He}{I} emission line. Together with the $\eta$~Car 
spectra, the spectra of the calibrators HD\,93030 and HD\,89682 are displayed. In addition, the telluric spectrum 
obtained at Kitt Peak with $R=40,000$ and a spectrally convolved telluric spectrum is shown, which matches the 
spectral resolution of the MR measurements ($R=1,500$). The figure illustrates that the forest of telluric lines 
forms a quasi-continuum which modulates the AMBER spectra.
}
  \label{fig:speccal}
\vspace*{-5mm}

\end{figure}

To obtain both an accurate wavelength calibration of the AMBER raw data and properly calibrated spectra of 
$\eta$~Car, we compared the AMBER raw spectra of $\eta$ Car as well as the calibrator stars L~Car, HD\,93030, 
and HD\,89682 with a $K$-band telluric spectrum recorded at the Kitt Peak Observatory with a spectral
resolution of 40,000. For the comparison with the AMBER spectrum, this telluric spectrum was spectrally 
convolved to match the spectral resolution of the AMBER measurements with high ($R\sim12,000$) and medium 
($R\sim1,500$) spectral resolution.

The result of the comparison is shown in Fig.~\ref{fig:speccal}. In the upper panel, the high spectral resolution 
AMBER spectra of $\eta$~Car and the calibrator L~Car are shown together with the telluric spectrum with $R=10,000$. 
>From the comparison with the telluric spectrum, we identified 7 prominent telluric absorption features in the L~Car 
spectrum, which are indicated by the dashed vertical lines. The strongest absorption line seen in the L~Car 
spectrum is not telluric, but can be identified as intrinsic Br$\gamma$ absorption in L Car. Therefore, to properly 
calibrate the $\eta$~Car spectrum with the L~Car spectrum, we had to interpolate the Br$\gamma$ line region in the 
L~Car spectrum before dividing the two spectra. From the spectral calibration shown in Fig.~\ref{fig:speccal}, we 
estimated a wavelength calibration error of the AMBER data $\Delta \lambda = 3\times10^{-4}\,\mu$m

The lower panel in Fig.~\ref{fig:speccal} shows the wavelength calibration of the medium spectral resolution data 
in the wavelength region around the \ion{He}{I} line. The figure contains the two $\eta$ Car MR spectra and the 
spectra of the two corresponding calibrator  stars, HD\,93030 and HD\,89682, as well as the telluric spectra with 
spectral resolutions of $R=40,000$ and $R=1,500$. As the telluric spectra reveal, there is a forest of telluric 
lines in the spectral region around the \ion{He}{I} line. As can be seen in Fig.~\ref{fig:overview.HeI.ALL}, the 
modulation of the continuum flux introduced by the telluric quasi-continuum cancels out completely when the 
$\eta$~Car spectra are divided by the corresponding calibrator spectra, which show no prominent intrinsic line 
features. Since there are no sharp spectral features in the 2.03--2.08$\,\mu$m region of either the calibrator or 
telluric spectras which could be used for the spectral calibration, we estimated a wavelength calibration error
$\Delta \lambda = 6\times10^{-4}\,\mu$m for the MR \ion{He}{I} data. On the other hand, for the MR data around 
the Br$\gamma$ line, we found $\Delta \lambda = 4\times10^{-4}\,\mu$m.

\section{Continuum uniform disk and Gauss diameter fits}\label{app:udgauss}

\begin{figure*}[t]
\centering
\includegraphics[width=53mm,angle=-90]{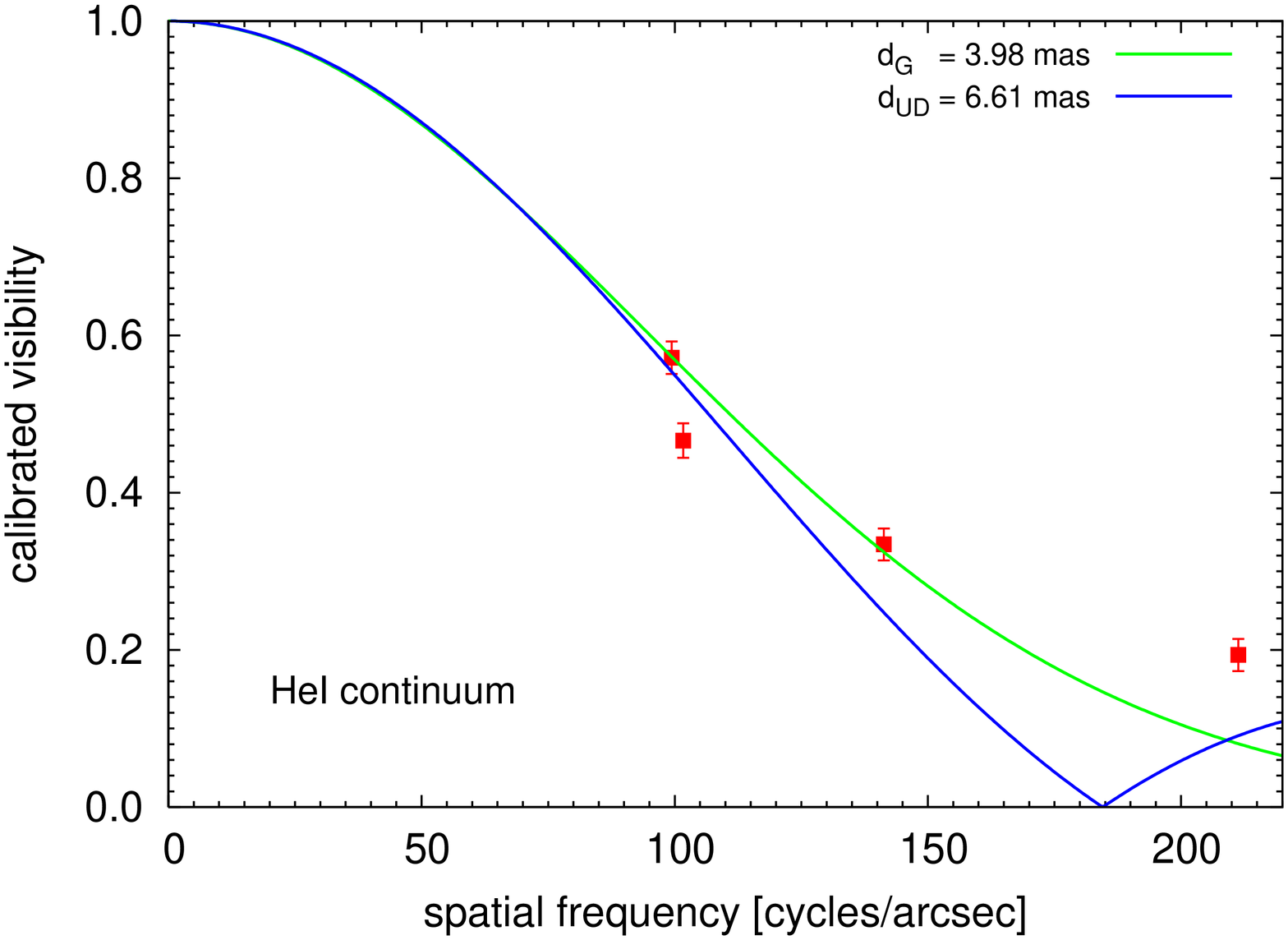}
\includegraphics[width=53mm,angle=-90]{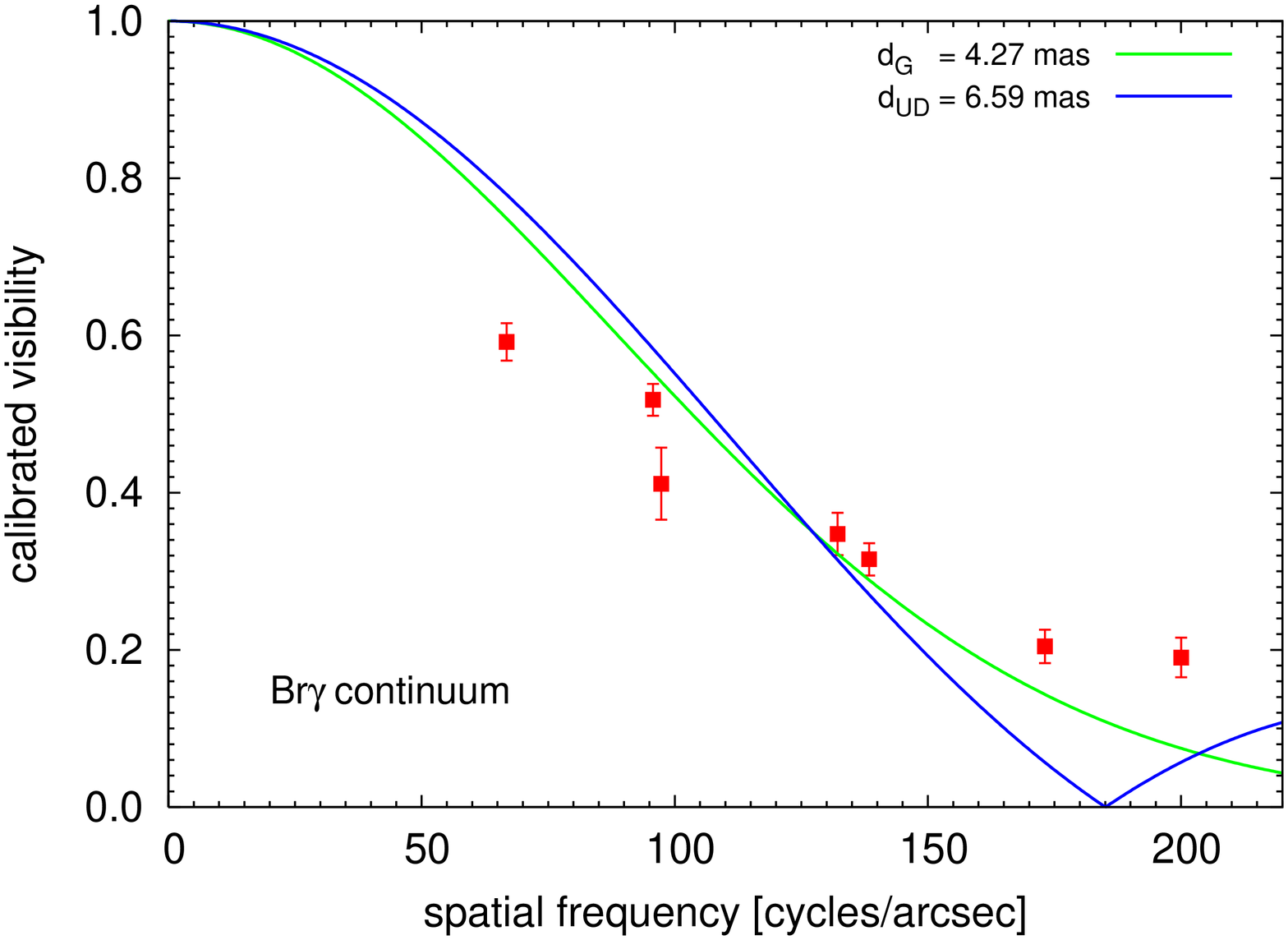}\\
\includegraphics[width=53mm,angle=-90]{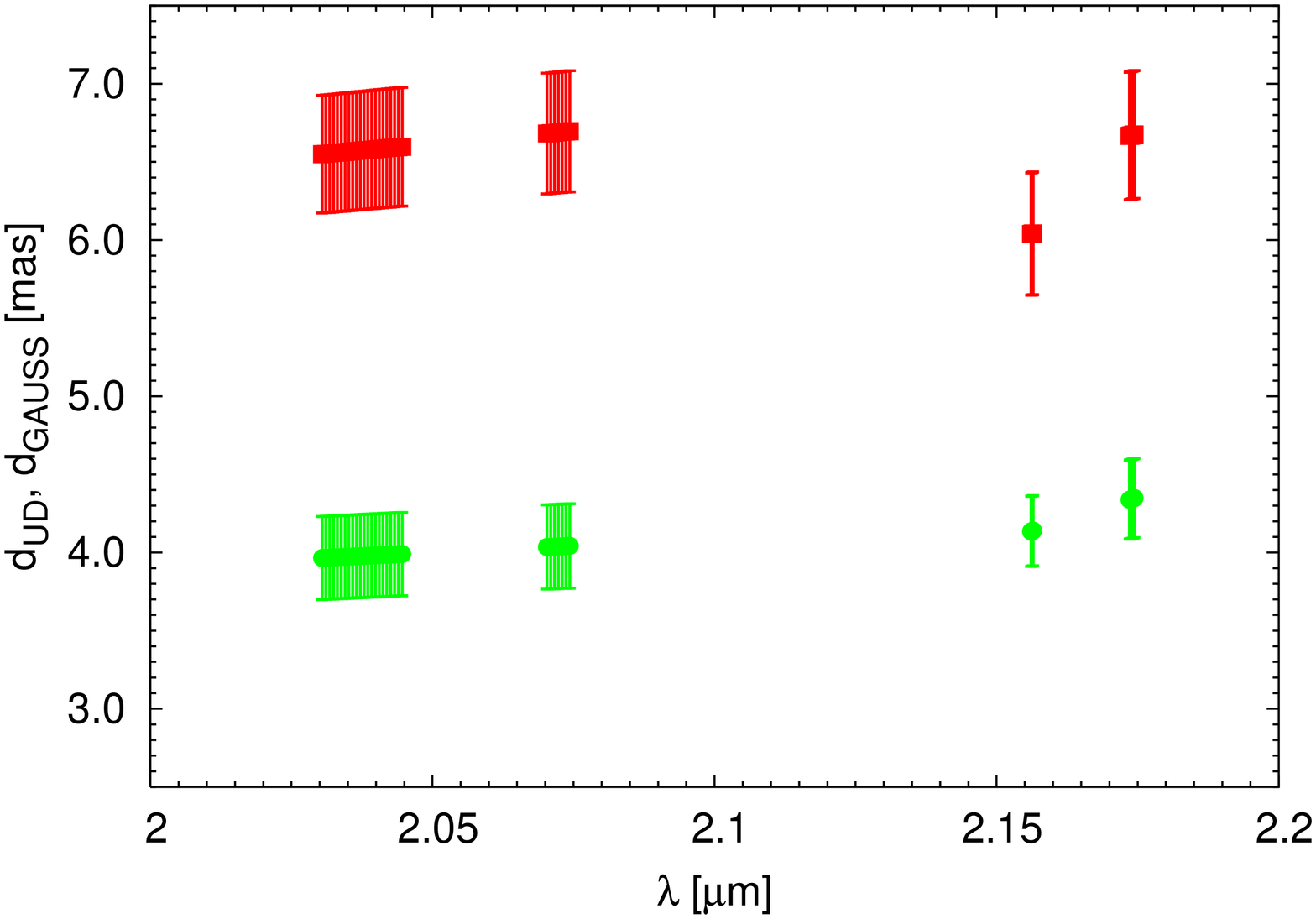}
\begin{minipage}[t]{75mm}
   \caption{\label{res:allvud}
   1-D visiblity fits of the AMBER continuum measurements. The upper two panels show uniform-disk and Gaussian fits 
   for the averaged continuum around the \ion{He}{I} and Br$\gamma$ lines, respectively. As both panels illustrate, 
   neither a uniform disk nor a Gaussian is a good representation of the AMBER measurements. The panel on the left 
   summarizes the fitting attempt for all continuum wavelength channels by displaying the fitted diameters of the  
   UD and GAUSS models as a function of wavelength.
}
\end{minipage}         
    \end{figure*}

For each spectral channel as well as for an averaged continuum, we performed 1-D fits to the visibility data using 
simple uniform disk (UD) and Gaussian models. In this step of the analysis, possible asymmetries were ignored and 
all visibility points at a given wavelength were fitted together, regardless of the position angle of the observations.
The results of these 1-D fits are illustrated in the two upper panels of Fig.~\ref{res:allvud} for the averaged 
continuum data in the wavelength ranges 2.03--2.08$\,\mu$m and 2.155--2.175$\,\mu$m, respectively. As the figure 
reveals, neither a uniform disk nor a single Gaussian provides a good fit to the continuum data. At least, this is 
true as long as no contamination by a fully resolved background component is taken into account.

\begin{figure}[h]
\centering
\includegraphics[width=57mm,angle=-90]{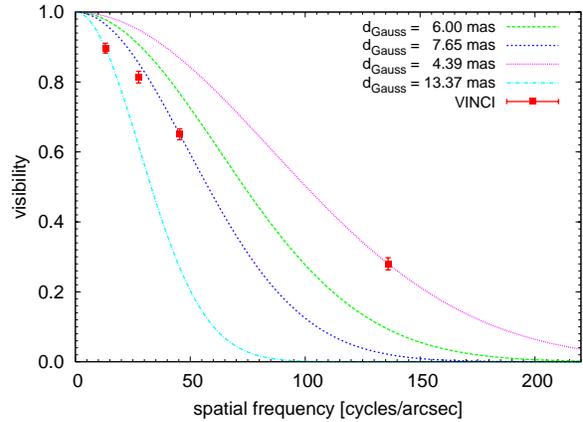}
\caption{\label{sect:app_vinci}
Dependence of the Gaussian FWHM diameter on the fit range. The figure shows the background-corrected
visibilities obtained with VLTI/VINCI \citep[see][]{van03} as well as Gaussian fits of (a) all four data 
points (long-dashed green line), (b) only the point with $q=45\,$ cycles/arcsec (short-dashed blue), (c) 
only the point corresponding to the longest baseline (dotted purple), and (d) only the point corresponding 
to the shortest baseline (dashed-dotted light blue). See the labels for the Gaussian FWHM diameters resulting 
from the different fits. The figure illustrates that the fitted diameter strongly depends on the spatial 
frequency range which is used to fit the data. The strong diameter variation (in this case, the diameter 
changes by a factor of $\sim3$) occurs since a Gaussian is not a good representation of the measured visibility 
function. See text for further discussion.
}
    \end{figure}

The wavelength dependence of the apparent size obtained from the UD and GAUSS model fits for the individual
spectral channels is shown in the lower left panel of Fig.~\ref{res:allvud}. This panel illustrates that the 
equivalent UD and GAUSS $K$-band diameters of $\eta$ Car derived from the AMBER data are $\sim$4 and $\sim$6.5 
mas, respectively.

It should be added here that a good fit of the AMBER data using, for instance, a Gaussian can indeed
be obtained when a certain amount of contamination due to a fully resolved background component is taken
into account \citep[see also][]{pet06}. To illustrate that, we performed Gaussian fits to the AMBER data, 
where we introduced such a fully resolved component as a free fitting parameter. We found that the best 
Gaussian fit is obtained with a FWHM diameter $d_{\rm Gauss} = 3.01$~mas and a $\sim29\%$ background 
contamination for the \ion{He}{I} continuum region and $d_{\rm Gauss} = 3.32$~mas and a $\sim30\%$ background 
contamination for the Br$\gamma$ continuum region. Thus, from this fit we would derive a background
contamination which is only $\sim50\%$ smaller than in the VINCI data. We think that such a large amount 
of background contamination is not very likely given the small AMBER fiber aperture (60~mas) of the 8.2~m 
telescopes. We think that the large amount of background contamination needed to find a reasonable Gauss 
fit just reflects the fact that a Gaussian is not appropriate to describe the observations. This is confirmed 
by the fact that for the fit of the radiative transfer model of \citet{hillier01}, no background component 
has to be taken into account to reproduce the AMBER measurements. 

We would like to note here that the Gaussian FWHM diameters of typically $\sim4$~mas found from the AMBER
measurements are not in contrast to the value $d_{\rm Gauss}\sim7$~mas found by \citet{van03} from VLTI/VINCI 
observations for the following reason: Since a Gaussian is not a good representation of both the VINCI and 
the AMBER visibilities, the diameter resulting from a Gaussian fit strongly depends on the fit range. 
This is illustrated in Fig.~\ref{sect:app_vinci} for the four VINCI measurements given in Fig.~1 of \citet{van03}.
As the figure shows, from a Gaussian fit of all four data points, $d_{\rm Gauss}=6.0$~mas is obtained.
If only the data point with $q=45\,$ cycles/arcsec is fitted (corresponding to a projected baseline length 
of $\sim24$~m), we get $d_{\rm Gauss}=7.65$~mas. This is in agreement with the values given in \citet{van03} 
for the elliptical Gaussian fit of the large number of VINCI measurements with a projected baseline of 24~m 
(see their Fig.~2). On the other hand, if we fit only the VINCI data point corresponding to the longest 
projected baseline ($q=136\,$ cycles/arcsec), a Gaussian fit provides $d_{\rm Gauss}=4.39$~mas (see 
Fig.~\ref{sect:app_vinci}), which is very close to the diameter we obtain from the AMBER measurements for 
$\lambda\sim2.174\,\mu$m ($d_{\rm Gauss}=4.35$~mas). This is not surprising since the spatial frequency of 
this VINCI data point agrees with the average spatial frequency of our AMBER observations ($q\sim$50--200 
cycles/arcsec). Thus, it can be concluded that good agreement between the Gaussian FWHM diameters derived 
from the AMBER and VINCI measurements is found if a comparable spatial frequency range is used for the fit.

\section{Visibility and differential phase of an emission line object}\label{app:dp}

We assume that the target's intensity distribution can be described by two components: the continuum spectrum 
$o_{\rm cont}(x,y;\lambda)$ and the emission line spectrum $o_{\rm line}(x,y;\lambda)$. In the part of the 
spectrum containing the emission line, both $o_{\rm cont}$ and $o_{\rm line}$ contribute to the total intensity 
distribution $o_{\rm tot}$. According to the van-Zittert-Zernike theorem, the Fourier transforms 
$O_{\rm cont}(B/\lambda)$ and $O_{\rm line}(B/\lambda)$ of $o_{\rm cont}(x,y;\lambda)$ and 
$o_{\rm line}(x,y;\lambda)$ are measured with an optical long baseline interferometer at wavelength $\lambda$ 
and projected baseline vector $B$. In the following, we assume that all Fourier spectra are normalized to 1 
at frequency zero. The complex Fourier spectrum $O_{\rm tot}(B/\lambda)$ of the intensity distribution 
$o_{\rm tot}$ measured at the emission line $\lambda_{\rm line}$ is given by
\begin{equation}
O_{\rm tot} = \frac{1}{F_{\rm cont}+F_{\rm line}} (F_{\rm cont}\cdot O_{\rm cont} + F_{\rm line}\cdot O_{\rm line}),
\label{app1:eq1}
\end{equation}
where $F_{\rm cont}$ and $F_{\rm line}$ are the fluxes of the continuum component $o_{\rm cont}$ and 
the line component $o_{\rm line}$, respectively. In the emission line, the total flux measured is 
$F_{\rm tot} := F_{\rm cont}+F_{\rm line}$.

>From the spectrally dispersed interferometric data, we can derive the differential phase, which is the difference 
of the Fourier phases of the continuum component $o_{\rm cont}$ and the total intensity $o_{\rm tot}$ in the 
emission line. The differential phase $\Phi'(B/\lambda_{\rm line})$ in the emission line at $\lambda_{\rm line}$ 
is given by
\begin{equation}
O_{\rm cont}\cdot O_{\rm tot}^* = V_{\rm cont}\cdot V_{\rm tot}\cdot e^{i\,\Phi'},
\label{app1:eq2}
\end{equation}
where $\Phi'(B/\lambda_{\rm line}) := \Phi_{\rm cont}(B/\lambda_{\rm line})-\Phi_{\rm tot}(B/\lambda_{\rm line})$.
$\Phi_{\rm cont}(B/\lambda_{\rm line})$ is the Fourier phase of the continuum component, and 
$\Phi_{\rm tot}(B/\lambda_{\rm line})$ denotes the Fourier phase of $o_{\rm tot}$ at the position of the emission 
line $\lambda_{\rm line}$. The asterisk $^*$ in this equation denotes conjugate complex operation.
$V_{\rm cont}(B/\lambda_{\rm line})$ describes the visibility of the continuum component at the position 
of the emission line $\lambda_{\rm line}$, and $V_{\rm tot}(B/\lambda_{\rm line})$ is the visibility measured 
at the position of the emission line $\lambda_{\rm line}$. Inserting Eq.~(\ref{app1:eq1}) into Eq.~(\ref{app1:eq2}) 
yields
\begin{displaymath}
V_{\rm cont}\cdot V_{\rm tot}\cdot e^{i\,\Phi'} \nonumber \\
\end{displaymath}
\begin{equation}
= \frac{V_{\rm cont}}{F_{\rm cont}+F_{\rm line}} \cdot ( F_{\rm cont} V_{\rm cont} + 
  F_{\rm line} V_{\rm line} e^{i\,\Delta\Phi} ),
\label{app1:eq3}
\end{equation}
where $\Delta\Phi(B/\lambda_{\rm line})$ denotes the difference of the Fourier phases of the continuum 
and line components; i.e.,
$\Delta\Phi(B/\lambda_{\rm line}) := \Phi_{\rm cont}(B/\lambda_{\rm line})-\Phi_{\rm line}(B/\lambda_{\rm line})$.
$\Phi_{\rm cont}(B/\lambda_{\rm line})$ and $\Phi_{\rm line}(B/\lambda_{\rm line})$ are the Fourier phases 
of the continuum and line components, respectively.

In the vector representation of complex numbers, the three quantities $F_{\rm cont} V_{\rm cont}$, 
$F_{\rm line} V_{\rm line},$ and $F_{\rm tot} V_{\rm tot}$ form a triangle with one corner placed 
at the center of the coordinate system. According to the law of cosines, the correlated flux of the 
line component is given by Eq.~(\ref{equ2}) (see Sect.~\ref{sect:res_brg}):
\begin{eqnarray}
& |F_{\rm line} V_{\rm line}|^2 & = |F_{\rm tot} V_{\rm tot}|^2 + |F_{\rm cont} V_{\rm cont}|^2 - \nonumber \\
& & - 2\cdot F_{\rm tot} V_{\rm tot} \cdot F_{\rm cont} V_{\rm cont} \cdot {\rm cos}(\Phi').
\label{app1:eq4}
\end{eqnarray}
Since the flux $F_{\rm line}$ can be calculated from the measured fluxes $F_{\rm cont}$ and $F_{\rm tot}$,
the visibility $V_{\rm line}$ of the line component can be derived using Eq.~(\ref{app1:eq4}).

Applying the law of sines to this triangle in the complex plane yields the differential phase 
$\Delta\Phi(B/\lambda_{\rm line})$, which is the difference between the Fourier phase 
$\Phi_{\rm cont}(B/\lambda_{\rm line})$ of the continuum component and the Fourier phase 
$\Phi_{\rm line}(B/\lambda_{\rm line})$ of the line component:
\begin{equation}
{\rm sin}(\Delta\Phi) = {\rm sin(\Phi')} \cdot \frac{|F_{\rm tot} V_{\rm tot}|}{|F_{\rm line} V_{\rm line}|},
\label{app1:eq5}
\end{equation}
where $\Phi'(B/\lambda_{\rm line})$ is the differential phase measured at the position of the 
emission line $\lambda_{\rm line}$.
\vspace*{3mm}

\end{document}